\documentclass[12pt]{article}
\usepackage{graphicx}
\usepackage{epsfig}
\usepackage{natbib}
\usepackage{amsmath}
\usepackage{amssymb}
\usepackage{booktabs}
\usepackage{rotating}
\usepackage{subfigure}
\usepackage[font=bf]{caption}
\usepackage[top=1in, bottom=1in, left=1in, right=1in]{geometry}

\newcommand{\nothing}[1]{{}}

\title{Double-jump stochastic volatility model for VIX: evidence from VVIX}
\author{\mbox{}\\ \mbox{}\\
\large Xin Zang%
\thanks{School of Mathematical Sciences, Peking University, Beijing 100871, PR China, USA; E-mail: xzang@pku.edu.cn.
}\\
Peking University\\ \mbox{}\\
\large Jun Ni%
\thanks{Department of Mathematics, Pennsylvania State University,
University Park, PA 16802, USA; E-mail: jzn132@psu.edu.
}\\
Penn State University\\ \mbox{}\\
\large Jing-Zhi Huang%
\thanks{Department of Finance, The Smeal College of Business, Pennsylvania State University,
University Park, PA 16802, USA; E-mail: jxh56@psu.edu.
}\\
Penn State University\\ \mbox{}\\
\large Lan Wu%
\thanks{Key Laboratory of Mathematical Economics and Quantitative Finance, Peking University, Beijing 100871, PR China; E-mail: lwu@pku.edu.cn.
}\\
Peking University\\ \mbox{}\\
\mbox{}\\
}
\date{\normalsize This version: \today}

\begin{document}
\maketitle

\clearpage

\vspace*{0.2in}

\setlength{\baselineskip}{8.5mm}

\begin{center}
{\Large Double-jump stochastic volatility model for VIX: evidence from VVIX}
\end{center}

\vspace*{0.5in}

\renewcommand{\arraystretch}{1.25}
\setlength{\baselineskip}{6mm}
\centerline{\bf \large Abstract}

\vspace*{0.2in}

The paper studies the continuous-time dynamics of VIX with stochastic
volatility and jumps in VIX and volatility. Built on the general parametric
affine model with stochastic volatility and jump in logarithm of VIX, we
derive a linear relation between the stochastic volatility factor and VVIX
index. We detect the existence of co-jump of VIX and VVIX and put forward a
double-jump stochastic volatility model for VIX through its joint property
with VVIX. With VVIX index as a proxy for the stochastic volatility, we use
MCMC method to estimate the dynamics of VIX. Comparing nested models on VIX,
we show the jump in VIX and the volatility factor is statistically
significant. The jump intensity is also state-dependent. We analyze the
impact of jump factor on the VIX dynamics.

\vspace{7mm}


\vspace{3mm}
\noindent{}

\thispagestyle{empty}

\pagebreak

\renewcommand{\arraystretch}{1.2}

\setlength{\baselineskip}{8.7mm}

\setcounter{page}{1}

\abovedisplayskip=10pt
\belowdisplayskip=10pt

\section{Introduction}

Modelling VIX index and its derivatives has been a hot topic among
researchers. As a measure for market's expectation of 30-day implied
volatility of S\&P500 index, VIX provides rich information for the
prediction of market's future trend. It can be seen as a compression of
information involved in S\&P500 options. Usually, the VIX and S\&P500 index
have an empirical negative correlation relationship, so VIX index is often
referred to as the fear index or the fear gauge. For more about VIX, see
e.g. \cite{Carr and Wu (2005)}.

Out of its importance, much attention has been focused on modelling the
dynamics of VIX directly. Earlier work tries using geometric Brownian
motion, square root diffusion or log-normal Ornstein-Uhlenbeck (OU) diffusion to
model VIX. Jumps in VIX are also added by some authors. Recently, a novel
parameterized stochastic volatility model for VIX is put forward by \cite%
{Mencia and Sentana (2013)} and \cite{Kaeck and Alexander (2013)}. They specify a new
process to model the volatility of VIX which may be correlated with VIX and
show its empirical advantage over the other traditional models. Both of them
also point out that the appearance of this stochastic volatility reduces the
impact of jump on VIX. However, the model specification of this novel
stochastic volatility diverges between them. \cite{Mencia and Sentana (2013)}
adopt a pure-jump OU process out of the analytical treatability while \cite%
{Kaeck and Alexander (2013)} characterize the volatility as a square root diffusion
which takes the correlation of the volatility and VIX into account. How to
specify and estimate this volatility factor and\ further interpret the
dynamics of VIX better is still an open problem.

In 2012 CBOE introduces a new volatility index named VVIX into market. Like
the role of VIX, VVIX measures the 30-day implied volatility of VIX index.
\cite{Huang and Shaliastovich (2014)} construct the realized volatility of VIX
index (i.e. realized volatility of volatility) and show that the VIX index
itself is not a good predictor for its realized volatility while VVIX serves
as a better candidate. We thus can infer from their empirical conclusion
that the VVIX index may provide some extra information about the volatility
of VIX beyond the VIX itself.

In this paper we mainly study the dynamics of VIX and
especially concentrate on modelling its stochastic volatility under the
physical measure via additional information provided from VVIX index. Based
on joint behavior of VIX and VVIX we put forward a double-jump stochastic
volatility model for logVIX and its volatility. We will show that under a
general affine assumption for the logarithm of VIX and its stochastic
volatility, the stochastic volatility and VVIX satisfy a linear
relationship, then the property of this stochastic volatility follows from
VVIX naturally. This relation can be seen as a benchmark to see whether the
estimated stochastic volatility factor is accurate enough and also provide
the empirical evidence for its model specification. From the historical data
of VIX and VVIX, we find both of them are mean-reverting. Furthermore,
through a formal test we find\ there exists evident co-jump between them.
Thus we conduct an empirical analysis of the double-jump stochastic
volatility model and its nested models using Markov-Chain-Monte-Carlo (MCMC)
method with historical data of VIX and VVIX. We provide evidence consistent with jumps in both VIX and the volatility and
that jump intensity is stochastic. We demonstrate the superiority
of our main model through residual analysis and simulation result.

The structure of this paper is as follows: Section \ref{section:literature
review} reviews the related literature. Section \ref{section:proxy} shows
the linear relationship between VVIX index and the stochastic volatility.
Section \ref{section:model specification} analyzes model specification and
sets ip our model. Section \ref{section:MCMC inference} gives our empirical
method. Section \ref{section:data description} describes the data of VIX and
VVIX that we use in this paper. Section \ref{section:empirical results}
summarizes the estimation results and provide empirical analysis. Section %
\ref{section:conclusion} concludes.

\section{Literature review}

\label{section:literature review}

To build a parameterized stochastic model for VIX, there are usually two
different starting points. One is first modelling a multi-factor stochastic
volatility process for S\&P500 index and then derive a calculating formula
for VIX index under this circumstance. For more about multi-factor model
setup, see e.g. \cite{Duffie et al. (2000)}, \cite{Gatheral (2008)},
\cite{Egloff et al. (2010)}, \cite{Cont and Kokholm (2013)} and \cite%
{Papanicolaou and Sircar (2014)}. From the model assumption, the final VIX may be a
combination of one or more factors (see e.g. \cite{Ait-Sahalia et al. (2014)}, \cite%
{Song and Xiu (2012)}, \cite{Luo and Zhang (2012)}, \cite{Lin and Chang (2009)}). The other method is
to model VIX index directly which often has mean-reverting property. In this
case, there are usually two ways to deal with the dynamics of VIX: affine
and non-affine (see e.g. \cite{Mencia and Sentana (2013)}, \cite%
{Kaeck and Alexander (2013)}, \cite{Goard and Mazur (2013)}). In the catalogue of
non-affine models, modelling logarithm of VIX directly is most popular and
has been proved empirically better than affine assumption of VIX.

Recently, modelling VIX with an additional stochastic volatility calls more
attention. Along the literature, the existence of stochastic volatility in
the dynamics of S\&P500 has been widely proved and accepted empirically.
Similarly, in terms of VIX index, the justification for stochastic
volatility is also tested and verified (see e.g. \cite{Wang and Daigler (2012)}, \cite%
{Huang and Shaliastovich (2014)}). Various authors has built VIX model with stochastic
volatility and found the model with stochastic volatility factor outperforms
the others without it. In \cite{Mencia and Sentana (2013)}, they make comparison
of different models for VIX with VIX, VIX futures and VIX options as the
data source. They use extended Kalman filter to estimate stochastic
volatility\ and conclude that modelling logarithm of VIX with stochastic
mean and stochastic volatility (named 'CTOUSV' in their paper) is the best
among all of the candidate models. In that paper, they first put forward the
stochastic volatility of volatility model of VIX and model the volatility
factor using pure jump OU process. \cite{Kaeck and Alexander (2013)} employ VIX
data of nearly 20 years to estimate the VIX models with and without
stochastic volatility using MCMC method. They model the volatility factor as
a square-root diffusion process and prove the stochastic volatility model
fits better for the VIX historical data. \cite{Barndorff-Nielsen and Veraart (2013)}
derive probabilistic properties of a class of stochastic volatility of
volatility models.

From \cite{Huang and Shaliastovich (2014)}, we conclude that using only VIX index to
infer the dynamics of the stochastic volatility factor of VIX is not a good
idea. At least some additional data set must be taken into consideration. In
terms of estimation, the data source matters. Estimation using various data
sources can produce distinct empirical results which show their different
information content (see e.g. \cite{Bardgett et al. (2013)}, \cite%
{Chung et al. (2011)}). In the framework of parameterized SDE model for
VIX, employing VIX and its derivatives (futures and options) has been
implemented before. However the formulas for VIX options are generally vey
complex and usually involve inverse Fourier transform which may increase the
calculation burden. Closed-form solutions are thus not easy to obtain for
VIX options. In that case, expansions is a good way (see e.g. \cite%
{Li (2013)}, \cite{Xiu (2014)}).\ The information from VIX options
are in essence equivalent to their implied volatility\ as other variables
like time to maturity\ and strikes are known. As a volatility index for VIX
options, VVIX just serves as the role of the implied volatility of VIX.
Using VVIX index to estimate dynamics of VIX is still a vacuum in the
literature and we will fill this gap in our paper.

\section{VVIX as a proxy for volatility of VIX}

\label{section:proxy}

\subsection{A motivation}

In \cite{Kaeck and Alexander (2013)}, they put forward a stochastic volatility of
volatility model for VIX. They assume the logarithm of VIX has a normal jump
while its stochastic volatility factor satisfies a square-root diffusion
model. Denote by $Y\left( t\right) $ the VIX index, we write this model
under $P$,%
\begin{eqnarray}
dY\left( t\right) &=&\kappa _{V}\left( \theta -Y\left( t\right) \right) dt+%
\sqrt{\omega \left( t\right) }dW_{Y}^{P}\left( t\right) +J_{Y}^{P}dN\left(
t\right)  \notag \\
d\omega \left( t\right) &=&\kappa _{\omega }^{P}\left( \varpi ^{P}-\omega
\left( t\right) \right) dt+\sigma _{\omega }\sqrt{\omega \left( t\right) }%
dW_{\omega }^{P}\left( t\right)  \label{Kaeck model}
\end{eqnarray}%
where $\left\langle dW_{Y}^{P}\left( t\right) ,dW_{\omega }^{P}\left(
t\right) \right\rangle =\rho dt$. $N\left( t\right) $ is Poisson process
with constant jump intensity $\lambda $,%
\begin{equation*}
J_{Y}^{P}\sim N\left( \mu _{y}^{JP},\left( \sigma _{y}^{J}\right) ^{2}\right)
\end{equation*}%
They use VIX data of 20 years as the input to estimate (\ref{Kaeck model})
and give the estimated latent stochastic volatility $\omega \left( t\right) $%
. A question arises naturally: how to guarantee and measure the accuracy of
the estimation of this unobserved spot volatility? \cite%
{Bardgett et al. (2013)} find that both of S\&P500 options and VIX options
contain some information that are unspanned by their counterpart. VIX is a
volatility index summarized from S\&P500 options while its diffusion part
must reflect the implied volatility of\ VIX options. So employing VIX index
as the only input to infer the dynamics of its stochastic volatility is
doubted. While VIX options can provide wanted extra information, the
computation burden rises quickly and sometimes complicated inverse Fourier
transform is needed out of the complex expression of the price of VIX
options. However, we will show in the next part that there exists a simple
linear relationship between the stochastic volatility factor and the VVIX
index which is compiled from a strip of VIX options mentioned above.

\subsection{Linear relationship between VVIX and volatility of VIX}

\label{linear relationship}

Following \cite{Mencia and Sentana (2013)} and \cite{Kaeck and Alexander (2013)}, in this
part we build model on the logarithm of VIX instead of VIX directly.\ We
show that if logVIX mean reverts to a constant central tendency with
stochastic volatility and jumps in logVIX and volatility, then there exists
a linear relationship between VVIX index and this stochastic volatility
factor of the logVIX. This relationship can provide a gauge to see whether a
stochastic volatility model for VIX can be reliable. The similar idea of
finding a proxy for some unobservable factor can also be found in \cite%
{Ait-Sahalia and Kimmel (2007)}, \cite{Duan and Yeh (2010)} and \cite{Ait-Sahalia et al. (2014)}. In this paper,
VIX index is the underlying asset so its dynamics are observed under $P$
measure.\ As VVIX is compiled from VIX options which is calculated under the
pricing measure $Q$, all of the derivation involving VVIX below will be
implemented under $Q$ measure.

Let $Y\left( t\right) =\log VIX\left( t\right) $ and assume that under $Q$, $%
Y\left( t\right) $ and $\omega \left( t\right) $ follow a general affine
jump diffusion model%
\begin{eqnarray}
dY\left( t\right) &=&\kappa _{V}\left( \theta -Y\left( t\right) \right) dt+%
\sqrt{\omega \left( t\right) }dW_{Y}^{Q}\left( t\right) +J_{Y}^{Q}dN\left(
t\right)  \notag \\
d\omega \left( t\right) &=&\left( \alpha _{\omega }-\kappa _{\omega
}^{Q}\omega \left( t\right) \right) dt+\sigma _{\omega }\sqrt{\omega \left(
t\right) }dW_{\omega }^{Q}\left( t\right) +J_{\omega }^{Q}dN\left( t\right)
\label{derivation setup}
\end{eqnarray}%
where we assume $\left\langle dW_{Y}^{Q}\left( t\right) ,dW_{\omega
}^{Q}\left( t\right) \right\rangle =\rho dt$. $N\left( t\right) $ is a
Poisson process with stochastic jump intensity $\lambda \left( t\right)
=\lambda _{0}+\lambda _{1}\omega \left( t\right) $ at time $t$ for analytical treatability. The jump for
VIX and its volatility factor are characterized by%
\begin{eqnarray*}
&&J_{Y}^{Q}\sim N\left( \mu _{y}^{J},\left( \sigma _{y}^{J}\right)
^{2}\right) \\
&&J_{\omega }^{Q}\text{ }\symbol{126}\text{ }N\left( \mu _{\omega
}^{J},\left( \sigma _{\omega }^{J}\right) ^{2}\right)
\end{eqnarray*}%

Similar to the idea that regarding VIX square as the expectation of quadratic variation of
the logarithm of the S\&P500 index under the pricing measure approximately (see, e.g. \cite{Ait-Sahalia et al. (2014)}), we set%
\begin{equation}
VVIX_{t,t+\tau }^{2}=\frac{1}{\tau }\left[ E_{t}^{Q}\left( \int_{t}^{t+\tau
}\omega \left( s\right) ds\right) +E_{t}^{Q}\left( \sum_{s\geq
0}\bigtriangleup Y^{2}\left( s\right) \right) \right]
\label{VVIX calculation definition}
\end{equation}%
With simple calculation from (\ref{derivation setup}) we obtain

\begin{eqnarray}
E_{t}^{Q}\left( \int_{t}^{t+\tau }\omega \left( s\right) ds\right)  &=&\frac{%
1-e^{-\left( \kappa _{\omega }^{Q}-\lambda _{1}\mu _{\omega }\right) \tau }}{%
\kappa _{\omega }^{Q}-\lambda _{1}\mu _{\omega }}\omega \left( t\right)
+\left( \tau -\frac{1-e^{-\left( \kappa _{\omega }^{Q}-\lambda _{1}\mu
_{\omega }\right) \tau }}{\kappa _{\omega }^{Q}-\lambda _{1}\mu _{\omega }}%
\right) \frac{\alpha _{\omega }+\lambda _{0}\mu _{\omega }}{\kappa _{\omega
}^{Q}-\lambda _{1}\mu _{\omega }}  \notag \\
&\triangleq &\alpha _{Q}\omega \left( t\right) +\beta _{Q}
\label{VVIX diffusion part}
\end{eqnarray}%
and%
\begin{eqnarray}
E_{t}^{Q}\left( \sum_{s\geq 0}\bigtriangleup Y^{2}\left( s\right) \right)
&=&\left( \left( \mu _{y}\right) ^{2}+\left( \sigma _{y}^{J}\right)
^{2}\right) E_{t}^{Q}\left( \int_{t}^{t+\tau }\left( \lambda _{0}+\lambda
_{1}\omega \left( s\right) \right) ds\right)   \notag \\
&=&\left( \left( \mu _{y}\right) ^{2}+\left( \sigma _{y}^{J}\right)
^{2}\right) \left( \lambda _{0}\tau +\lambda _{1}\beta _{Q}+\lambda
_{1}\alpha _{Q}\omega \left( t\right) \right)   \label{VVIX jump part}
\end{eqnarray}%
Combine (\ref{VVIX calculation definition}), (\ref{VVIX diffusion part}) and
(\ref{VVIX jump part}), we finally have
\begin{eqnarray}
VVIX_{t,t+\tau }^{2} &=&\frac{1}{\tau }\left[ \alpha _{Q}\omega \left(
t\right) +\beta _{Q}+\left( \left( \mu _{y}\right) ^{2}+\left( \sigma
_{y}^{J}\right) ^{2}\right) \left( \lambda _{0}\tau +\lambda _{1}\beta
_{Q}+\lambda _{1}\alpha _{Q}\omega \left( t\right) \right) \right]  \notag \\
&=&\frac{1}{\tau }\left[ \left( \beta _{Q}+\left( \left( \mu _{y}\right)
^{2}+\left( \sigma _{y}^{J}\right) ^{2}\right) \left( \lambda _{0}\tau
+\lambda _{1}\beta _{Q}\right) \right) +\left( 1+\lambda _{1}\left( \left(
\mu _{y}\right) ^{2}+\left( \sigma _{y}^{J}\right) ^{2}\right) \right)
\alpha _{Q}\omega \left( t\right) \right]  \notag \\
&\triangleq &A\left( \tau \right) +B\left( \tau \right) \omega \left(
t\right)  \label{VVIX formula}
\end{eqnarray}

Relationship (\ref{VVIX formula}) can be seen as a benchmark for the
estimated volatility factor. The dynamics of VVIX can reflect the property
of $\omega \left( t\right) $ more directly than the indirect impact of VIX
option. It can provide more intuitive empirical evidence for the model
specification for $\omega \left( t\right) $ which will be seen in Section %
\ref{section:model specification}.

\subsection{Examination using the benchmark}

If we define a suitable set of risk premia specification\ for this model to
guarantee that its $Q$-counterpart remains the same structure as that under $%
P$,\ then the linear relation in Section \ref{linear relationship} holds.\
We want to use VVIX index as a benchmark (let $J_{\omega }^{Q}=0$, $\lambda
_{1}=0$) to judge whether the estimation for $\omega \left( t\right) $ from
only VIX index is reliable and reflect the real evolution of the market. As
the VVIX data only starts from 2007, so in this part, we use VIX data from
Jan 2007 to Sep 2014 to estimate this model again using the same method as
\cite{Kaeck and Alexander (2013)}. We plot the estimated volatility factor $\omega
\left( t\right) $ and VVIX time series of the same period in Figure \ref%
{Spot Volatility-VVIX}. The correlation between this posterior volatility
and VVIX index is only 0.4193. Although some of the peaks of estimated $%
\omega \left( t\right) $ coincide with VVIX, more inconsistence between them
appears. This indicates that when we use VIX index as the only data source
to sample latent variable $\omega \left( t\right) $, it could only provide
limited information about the dynamics of its stochastic volatility and the
stochastic volatility is unspanned by the VIX in some sense. To obtain more
accurate $\omega \left( t\right) $, the relationship between it and the VVIX
index can be utilized.

\section{Model specification and setup}

\label{section:model specification}

\subsection{Model specification}

The log-normal Ornstein-Uhlenbeck model is put forward by \cite%
{Detemple and Osakwe (2000)}. Ever since, modelling the logarithm of VIX or VIX
futures is considered in \cite{Psychoyios et al. (2010)} and \cite{Huskaj and Nossman (2013)}%
. \cite{Mencia and Sentana (2013)} and \cite{Kaeck and Alexander (2013)} compare and
examine different model specification for VIX dynamics. Both of them
conclude that\ the setup for modelling logVIX as an affine jump process is
superior to modelling VIX directly which is consistent among all of the
model specifications in more detail. So in our model, we also study the
affine property of logVIX.

Since there exists such linear relationship between VVIX and $\omega \left(
t\right) $, the VVIX index can be seen as a proxy for this unobservable
variable. The joint modelling of VIX index and its stochastic volatility is
thus\ equivalent to the joint modelling for VIX index and VVIX index. In
this sense, the model should reflect some of their joint property.

Both VIX and VVIX have the mean-reverting property. For VIX, it may mean
revert to a constant or stochastic central tendency. In \cite%
{Mencia and Sentana (2013)}, they make both assumptions and examine the model
performance respectively. As their data source consists of VIX, VIX futures
and VIX options, the stochastic central tendency of VIX models performs
better. In fact the specification of the central tendency of VIX is mainly
characterized by the information from VIX futures while the VIX options play
a relatively light role. However, from the derivation in Section \ref{linear
relationship} we know that the expression of VVIX index is irrelevant of the
drift part of VIX. In fact, this is consistent with its stochastic
volatility role. As our paper mainly concerns about the impact of VVIX data
for the estimation, we make a simple assumption that the VIX mean reverts to
a constant central tendency. As VVIX has the similar empirical property, we
also assume the stochastic volatility of VIX has a constant central tendency.

From historical daily data of VIX and VVIX from Jan 2007 to Nov 2014, we
observe that there exists evident co-jump between\ the two index, no matter
positive or negative jump happened. To make a formal test to verify this
phenomenon, we adapt the method in \cite{Bollerslev et al. (2008)} to the lower
sampling frequency (see also \cite{Gilder (2009)} for the practice of
this method for daily data). The testing procedure is divided into two
steps: first, we show there exist jumps in both VIX and VVIX and second, the
VIX and VVIX have common jumps.

To carry on the first step, we assume a process $X$ (to be VIX or VVIX) is
observed in $\left[ 0,T\right] $ at daily times $t=0,1,\ldots ,T$ and denote
the time series by $X_{t},t=1,2,\ldots ,T$. The return process $%
r_{t}=X_{t}-X_{t-1},t=1,2,\ldots ,T$ is also defined.\ We compute the $n$%
-day rolling sample estimates of realized volatility,%
\begin{equation}
RV_{t}=\sum\nolimits_{k=0}^{n}r_{t-k}^{2}  \label{RV}
\end{equation}%
and bipower variation%
\begin{equation}
BV_{t}=\frac{\pi }{2}\sum\nolimits_{k=0}^{n-1}\left\vert r_{t-k}\right\vert
\left\vert r_{t-k-1}\right\vert  \label{BV}
\end{equation}%
The relative contribution measure%
\begin{equation}
RJ_{t}=\frac{RV_{t}-BV_{t}}{RV_{t}}  \label{RJ}
\end{equation}%
follows from (\ref{RV}) and (\ref{BV}) immediately.\ The tripower quarticity
for daily changes is defined by%
\begin{equation}
TP_{t}=\mu _{4/3}^{-3}\frac{n^{2}}{n-2}\sum\nolimits_{k=0}^{n-1}\left\vert
r_{t-k}\right\vert ^{4/3}\left\vert r_{t-k-1}\right\vert ^{4/3}\left\vert
r_{t-k-2}\right\vert ^{4/3}  \label{TP}
\end{equation}%
where $\mu _{4/3}=2^{2/3}\Gamma \left( \frac{7}{6}\right) \Gamma \left(
\frac{1}{2}\right) $. Finally, the statistic%
\begin{equation}
z_{t}=\frac{RJ_{t}}{\sqrt{\left[ \left( \pi /2\right) ^{2}+\pi -5\right]
\frac{1}{n}\max \left( 1,\frac{TP_{t}}{BV_{t}^{2}}\right) }}  \label{Zt}
\end{equation}%
is constructed using (\ref{BV}), (\ref{RJ}) and (\ref{TP}) to test whether a
jump occurs at day $t$. We reject the null hypothesis of no jumps at $\alpha
\%$ confidence level if $\left\vert z_{t}\right\vert >\Phi _{1-\alpha
/2}^{-1}$ where $\Phi $ is the cumulative normal distribution for a given
day $t$.

To implement the second step, denote the VIX and VVIX by $X^{1}$ and $X^{2}$%
. Assume VIX and VVIX index are observed in $\left[ 0,T\right] $ at daily
times $t=1,2,\ldots ,T$, the time series are thus $X_{t}^{i},t=1,2,\ldots
,T,i=1,2$ respectively. Given the return processes $%
r_{t}^{i}=X_{t}^{i}-X_{t-1}^{i},t=1,2,\ldots ,T,i=1,2$, we calculate the
contemporaneous correlation%
\begin{equation*}
cp_{t}=\sum\nolimits_{k=0}^{n-1}r_{t-k}^{1}r_{t-k}^{2}
\end{equation*}%
and study the studentized statistic%
\begin{equation}
z_{cp,t}=\frac{cp_{t}-\overline{cp}}{s_{cp}}  \label{Zcpt}
\end{equation}%
where%
\begin{equation*}
\overline{cp}=\frac{1}{T-\left( n-1\right) }\sum\nolimits_{t=n}^{T}cp_{t}
\end{equation*}%
and%
\begin{equation*}
s_{cp}=\left[ \frac{1}{T-\left( n-1\right) }\sum\nolimits_{t=n}^{T}\left(
cp_{t}-\overline{cp}\right) ^{2}\right] ^{1/2}
\end{equation*}%
at time $t$. We reject the null hypothesis of no common jumps at $\alpha \%$
confidence level if $\left\vert z_{cp,t}\right\vert >\Phi _{1-\alpha
/2}^{-1} $ where $\Phi $ is the cumulative normal distribution for a given
day $t$.

Employing the methods given above, we test the jump behavior of VIX and VVIX
from January 3, 2007 to November 26, 2014. Given the 5\% significant level,
222 days for VIX and 141 days for VVIX\ out of 1939 days indicate the
significant jump for the first step. In the second step, 131 days call for
co-jump. Thus the specification for co-jump in VIX and VVIX is justified and
this phenomenon provides an important foundation for our model setup.

\subsection{Basic model}

This demonstrates that in addition to the diffusion part,\ we should assume
jump in both $Y\left( t\right) $ and $\omega \left( t\right) $ and the
jumps\ should be dominated by\ a single Poisson process. The jump intensity
may be constant or state-dependent on the affine factor. In this paper we
assume it is affected by $\omega \left( t\right) $. The assumption for
constant or stochastic jump intensity will be examined below. As both
positive and negative jumps appear, we make the normal distribution
assumption for the jump size. For logVIX this may be a sensible assumption.
While for the square root diffusion plus a jump for $\omega \left( t\right) $%
, as jump is a rare event for the historical path, the assumption is also
acceptable. For more previous work on jumps in volatility, we refer to \cite%
{Duffie et al. (2000)}, \cite{Eraker et al. (2003)}, \cite{Eraker (2004)},
\cite{Todorov and Tauchen (2011)} and \cite{Amengual and Xiu (2014)}.

We thus assume that under $Q$,%
\begin{eqnarray*}
dY\left( t\right) &=&\kappa _{V}\left( \theta -Y\left( t\right) \right) dt+%
\sqrt{\omega \left( t\right) }dW_{Y}^{Q}\left( t\right) +J_{Y}^{Q}dN\left(
t\right) \\
d\omega \left( t\right) &=&\left( \alpha _{\omega }-\kappa _{\omega
}^{Q}\omega \left( t\right) \right) dt+\sigma _{\omega }\sqrt{\omega \left(
t\right) }dW_{\omega }^{Q}\left( t\right) +J_{\omega }^{Q}dN\left( t\right)
\end{eqnarray*}%
where $\left\langle dW_{Y}^{Q}\left( t\right) ,dW_{\omega }^{Q}\left(
t\right) \right\rangle =\rho dt$ and $N\left( t\right) $ is a Poisson
process with stochastic jump intensity $\lambda \left( t\right) =\lambda
_{0}+\lambda _{1}\omega \left( t\right) $ at time $t$.%
\begin{eqnarray*}
&&J_{Y}^{Q}\sim N\left( \mu _{y}^{J},\left( \sigma _{y}^{J}\right)
^{2}\right) \\
&&J_{\omega }^{Q}\text{ }\symbol{126}\text{ }N\left( \mu _{\omega
}^{J},\left( \sigma _{\omega }^{J}\right) ^{2}\right)
\end{eqnarray*}%
We specify the risks of price between $Q$ and $P$ about Brownian motions as%
\begin{eqnarray*}
dW_{Y}^{Q}\left( t\right) &=&dW_{Y}^{P}\left( t\right) -\varsigma _{V}\sqrt{%
\omega \left( t\right) }dt \\
dW_{\omega }^{Q}\left( t\right) &=&dW_{\omega }^{P}\left( t\right)
-\varsigma _{\omega }\sqrt{\omega \left( t\right) }dt
\end{eqnarray*}%
then under $P$,%
\begin{eqnarray}
dY\left( t\right) &=&\left[ \kappa _{V}\left( \theta -Y\left( t\right)
\right) -\varsigma _{V}\omega \left( t\right) \right] dt+\sqrt{\omega \left(
t\right) }dW_{Y}^{P}\left( t\right) +J_{Y}^{P}dN\left( t\right)  \notag \\
d\omega \left( t\right) &=&\left( \alpha _{\omega }-\kappa _{\omega
}^{P}\omega \left( t\right) \right) dt+\sigma _{\omega }\sqrt{\omega \left(
t\right) }dW_{\omega }^{P}\left( t\right) +J_{\omega }^{P}N\left( t\right)
\label{P model}
\end{eqnarray}%
where $\left\langle dW_{Y}^{P}\left( t\right) ,dW_{\omega }^{P}\left(
t\right) \right\rangle =\rho dt$ and $N\left( t\right) $ is a Poisson
process with stochastic jump intensity $\lambda \left( t\right) =\lambda
_{0}+\lambda _{1}\omega \left( t\right) $ at time $t$. $\kappa _{\omega
}^{P}=\kappa _{\omega }^{Q}+\varsigma _{\omega }\sigma _{\omega }$ is the
speed of mean reversion under $P$. The jump sizes are characterized by
\begin{eqnarray*}
J_{Y}^{P} &\sim &N\left( \mu _{y}^{JP},\left( \sigma _{y}^{J}\right)
^{2}\right) \\
J_{\omega }^{P} &\sim &N\left( \mu _{\omega }^{JP},\left( \sigma _{\omega
}^{J}\right) ^{2}\right)
\end{eqnarray*}

The parameter set under $P$ is denoted by
\begin{equation*}
\Theta _{P}=\left\{ \kappa _{V},\varsigma _{V},\theta ,\kappa _{\omega
}^{P},\mu _{y}^{JP},\mu _{\omega }^{JP},\sigma _{\omega }^{J},\rho ,\sigma
_{\omega }\right\}
\end{equation*}%
which the parameter set under $Q$ is summarized as%
\begin{equation*}
\Theta _{M}=\left\{ \alpha _{\omega },\kappa _{\omega }^{Q},\lambda
_{0},\lambda _{1},\mu _{y},\mu _{\omega },\sigma _{y}^{J}\right\}
\end{equation*}

We also assume that there exists a pricing error for VVIX from our theoretical model, s.t.%
\begin{equation}
VVIX_{t,t+\tau }^{2}=A\left( \tau \right) +B\left( \tau \right) \omega
\left( t\right) +\varepsilon  \label{linear relationship setup}
\end{equation}%
and%
\begin{equation*}
\varepsilon \sim N\left( 0,\sigma _{P}^{2}\right)
\end{equation*}%
we need to estimate%
\begin{equation*}
\Theta _{E}=\left\{ \sigma _{P}\right\}
\end{equation*}

We call the general model (\ref{P model}) the SVJJ-S model (stochastic $%
\lambda $). If we let $\lambda _{1}=0$, it reduces to the SVJJ-C model
(constant $\lambda $) model. If we further let $J_{\omega }^{P}\left(
J_{\omega }^{Q}\right) =0$, it collapses to the SVJ-C model (constant $%
\lambda $) model. Finally, when there are no jumps, i.e., $J_{y}=J_{\omega
}=0$, we call it SV model. We want to examine these models using the real
market historical data of VIX and VVIX to see: 1, whether adding the jump
into VIX and $\omega \left( t\right) $ can improve the VIX model
significantly; 2, whether the jump intensity is constant or stochastic.

\section{Model inference with VIX and VVIX}

\label{section:MCMC inference}

In this part, we use VIX and VVIX index data from January 3, 2007 to
November 26, 2014 to estimate the models. In total we have 1991 daily
observations for VIX and VVIX index respectively. We adopt MCMC method as
the estimation method. Compared with maximum-likelihood estimation (MLE),
generalized method of moments (GMM) and some other methods, MCMC has two
advantages that adapts to our aim. First, not only does MCMC estimate the
unknown parameters, it can also provide posterior estimated latent variables
such as stochastic volatility, jump times and jump sizes. These variables
are fundamental and essential for subsequent empirical analysis and model
comparison. Second, MCMC is very efficient for implementation. For more
details about applications of MCMC method in finance, we refer to \cite%
{Johannes and Polson (2003)}, \cite{Eraker et al. (2003)} and \cite{Amengual and Xiu (2012)}.

Denote by the parameters set by $\mathbf{\Theta }=\left( \Theta _{P},\Theta
_{M},\Theta _{E}\right) $, the latent variables by $\mathbf{Z}$ and the
observed data by $\mathbf{Y}=\left( VIX,VVIX\right) $, for some model $M$ we
are interested in the joint posterior of parameters and latent variables
given data:%
\begin{equation*}
p\left( \mathbf{\Theta ,Z}|\mathbf{Y},M\right) \varpropto p\left( \mathbf{%
Y|\Theta ,Z},M\right) \cdot p\left( \mathbf{\Theta ,Z}|M\right)
\end{equation*}

We assume the market data are observed daily. Let the time interval $\Delta
=1/252$ be one day, assume we have $T$ observations $Y_{i\Delta },0\leq
i\leq T+1$ for logarithm of VIX.\ A time discretization of the dynamics (\ref%
{P model}) with time interval $\Delta $ gives%
\begin{eqnarray}
Y_{i\Delta }-Y_{\left( i-1\right) \Delta } &=&\left( \kappa _{V}\theta
-\kappa _{V}Y_{\left( i-1\right) \Delta }-\varsigma _{V}\omega _{\left(
i-1\right) \Delta }\right) \Delta +\sqrt{\omega _{\left( i-1\right) \Delta
}\Delta }\epsilon _{i\Delta }^{y}+j_{i\Delta }^{y}n_{i\Delta }  \notag \\
\omega _{i\Delta }-\omega _{\left( i-1\right) \Delta } &=&\left( \alpha
_{\omega }-\kappa _{\omega }^{P}\omega _{\left( i-1\right) \Delta }\right)
\Delta +\sigma _{\omega }\sqrt{\omega _{\left( i-1\right) \Delta }\Delta }%
\epsilon _{i\Delta }^{\omega }+j_{i\Delta }^{\omega }n_{i\Delta }
\label{original discretization}
\end{eqnarray}%
where $\epsilon _{i\Delta }^{y}$ and $\epsilon _{i\Delta }^{\omega }$ are
correlated Normal variables with correlation $\rho $, $j_{i\Delta }^{y}$ and
$j_{i\Delta }^{\omega }$ are normal with different parameters.

Denote by $\widetilde{Y}_{i\Delta }=Y_{i\Delta }-j_{i\Delta }^{y}n_{i\Delta
} $ for $2\leq i\leq T+1$\ and $\widetilde{\omega }_{i\Delta }=\omega
_{i\Delta }-j_{i\Delta }^{\omega }n_{i\Delta }$ for $2\leq i\leq T$, then we
transform from (\ref{original discretization}) to the jump-adjusted processes%
\begin{eqnarray}
\widetilde{Y}_{i\Delta } &=&a_{0}+a_{1}Y_{\left( i-1\right) \Delta
}+a_{2}\omega _{\left( i-1\right) \Delta }+\sqrt{\omega _{\left( i-1\right)
\Delta }\Delta }\epsilon _{i\Delta }^{y}  \notag \\
\widetilde{\omega }_{i\Delta } &=&c_{0}+c_{1}\omega _{\left( i-1\right)
\Delta }+\sigma _{\omega }\sqrt{\omega _{\left( i-1\right) \Delta }\Delta }%
\epsilon _{i\Delta }^{\omega }  \label{jump-adjusted processes}
\end{eqnarray}%
where $a_{0}=\kappa _{V}\theta \Delta $, $a_{1}=1-\kappa _{V}\Delta $, $%
a_{2}=-\varsigma _{V}\Delta $, $c_{0}=\alpha _{\omega }\Delta $, $%
c_{1}=1-\kappa _{\omega }^{P}\Delta $. In this part, we will apply $%
Y_{i\Delta },0\leq i\leq T+1$ to estimate latent variables%
\begin{eqnarray*}
\omega _{i\Delta },\text{ }1 &\leq &i\leq T \\
n_{i\Delta },j_{i\Delta }^{y}\text{ and }j_{i\Delta }^{\omega },\text{ }2
&\leq &i\leq T+1
\end{eqnarray*}

As the joint posterior distribution $p\left( \mathbf{\Theta ,Z|M}\right) $
are not known in closed-form, the MCMC algorithm samples these parameters
and latent variables sequentially from posterior conditional distributions
as follows:%
\begin{eqnarray*}
\text{spot volatility}\text{: } &&p\left( \omega _{i\Delta }^{\left(
g\right) }|\omega _{<i\Delta }^{\left( g\right) },\omega _{>i\Delta
}^{\left( g-1\right) },n_{i\Delta }^{\left( g-1\right) },j_{i\Delta
}^{y\left( g-1\right) },j_{i\Delta }^{\omega \left( g-1\right) },\Theta
^{\left( g-1\right) },Y\right) \\
\text{jump time} &\text{:}&p\left( n_{i\Delta }^{\left( g\right) }|\omega
_{i\Delta }^{\left( g\right) },j_{i\Delta }^{y\left( g-1\right) },j_{i\Delta
}^{\omega \left( g-1\right) },\Theta ^{\left( g-1\right) },Y\right) \\
\text{jump size in VIX} &\text{:}&p\left( j_{i\Delta }^{y\left( g\right)
}|n_{i\Delta }^{\left( g\right) },\omega _{i\Delta }^{\left( g\right)
},j_{i\Delta }^{\omega \left( g-1\right) },\Theta ^{\left( g-1\right)
},Y\right) \\
\text{jump size in volatility} &\text{:}&p\left( j_{i\Delta }^{\omega \left(
g\right) }|n_{i\Delta }^{\left( g\right) },\omega _{i\Delta }^{\left(
g\right) },j_{i\Delta }^{y\left( g\right) },\Theta ^{\left( g-1\right)
},Y\right) \\
\text{parameters} &\text{:}&p\left( \Theta ^{\left( g\right) }|n_{i\Delta
}^{\left( g\right) },\omega _{i\Delta }^{\left( g\right) },j_{i\Delta
}^{y\left( g\right) },j_{i\Delta }^{\omega \left( g\right) },\Theta ^{\left(
g-1\right) },Y\right)
\end{eqnarray*}%
where $g$ represents the iteration times. In this paper, we sample 5000
times and discard the first 2000 samples.

\subsection{Estimation Strategy}

In this part we consider the sampling method for the latent variables and
parameters. We will discuss the corresponding algorithms for the stochastic
volatility $\omega _{t}$, the $Q$-parameters $\Theta _{M}$ and the pricing
error parameter $\Theta _{E}$. For jump times, jump sizes and parameters in $%
\Theta _{P}$, the sampling methods are standard and\ Appendix gives
detailed algorithm.

Sampling the stochastic volatility $\omega _{t}$ should take the information
from both VIX and VVIX into consideration. Utilizing the linear relationship
in (\ref{linear relationship setup}), we use random-walk metropolis method
to sample $\omega _{t}$. Let $\omega _{\left( -i\right) }^{\left( g-1\right)
}=\left( \omega _{1\Delta }^{\left( g\right) },\cdots ,\omega _{\left(
i-1\right) \Delta }^{\left( g\right) },\omega _{\left( i+1\right) \Delta
}^{\left( g-1\right) },\cdots ,\omega _{T\Delta }^{\left( g-1\right)
}\right) $ where the index $\left( g\right) $ represents the iteration
times, we specify the full conditional density as
\begin{eqnarray*}
&&p_{i}\left( \omega _{i\Delta }^{\left( g\right) }|\omega _{\left(
-i\right) }^{\left( g-1\right) },n_{i\Delta }^{\left( g-1\right)
},j_{i\Delta }^{y\left( g-1\right) },j_{i\Delta }^{\omega \left( g-1\right)
},\Theta ^{\left( g-1\right) },Y\right)  \\
&\propto &\frac{1}{\omega _{i\Delta }^{\left( g\right) }}\exp \left[ -\frac{%
\left( C_{i\Delta }^{2}+D_{i\Delta }^{2}-2\rho C_{i\Delta }D_{i\Delta
}\right) }{2\left( 1-\rho ^{2}\right) }\right] \exp \left[ -\frac{\left(
C_{\left( i+1\right) \Delta }^{2}+D_{\left( i+1\right) \Delta }^{2}-2\rho
C_{\left( i+1\right) \Delta }D_{\left( i+1\right) \Delta }\right) }{2\left(
1-\rho ^{2}\right) }\right]  \\
&&\cdot \exp \left( -\frac{\left( VVIX_{i\Delta }^{2}-A\left( \tau \right)
-B\left( \tau \right) \omega _{i\Delta }^{\left( g\right) }\right) ^{2}}{%
2\sigma _{P}^{2}}\right)
\end{eqnarray*}%
where%
\begin{eqnarray*}
C_{i\Delta } &=&\frac{Y_{i\Delta }-j_{i\Delta }^{y\left( g-1\right)
}n_{i\Delta }^{\left( g-1\right) }-a_{0}-a_{1}Y_{\left( i-1\right) \Delta
}-a_{2}\omega _{\left( i-1\right) \Delta }^{\left( g\right) }}{\sqrt{\omega
_{\left( i-1\right) \Delta }^{\left( g\right) }\Delta }} \\
D_{i\Delta } &=&\frac{\omega _{i\Delta }^{\left( g\right) }-j_{i\Delta
}^{\omega \left( g-1\right) }n_{i\Delta }^{\left( g-1\right)
}-c_{0}-c_{1}\omega _{\left( i-1\right) \Delta }^{\left( g\right) }}{\sigma
_{\omega }\sqrt{\omega _{\left( i-1\right) \Delta }^{\left( g\right) }\Delta
}}
\end{eqnarray*}%
and%
\begin{eqnarray*}
C_{\left( i+1\right) \Delta } &=&\frac{Y_{\left( i+1\right) \Delta
}-j_{\left( i+1\right) \Delta }^{y\left( g-1\right) }n_{\left( i+1\right)
\Delta }^{\left( g-1\right) }-a_{0}-a_{1}Y_{\left( i-1\right) \Delta
}-a_{2}\omega _{i\Delta }^{\left( g\right) }}{\sqrt{\omega _{i\Delta
}^{\left( g\right) }\Delta }} \\
D_{\left( i+1\right) \Delta } &=&\frac{\omega _{\left( i+1\right) \Delta
}^{\left( g-1\right) }-j_{\left( i+1\right) \Delta }^{\omega \left(
g-1\right) }n_{\left( i+1\right) \Delta }^{\left( g-1\right)
}-c_{0}-c_{1}\omega _{i\Delta }^{\left( g\right) }}{\sigma _{\omega }\sqrt{%
\omega _{i\Delta }^{\left( g\right) }\Delta }}
\end{eqnarray*}%
for $2\leq i\leq T-1$. The case for $i=1$ and $i=T$ follows similarly. Note
that this target density contains information from both VIX and VVIX.

The $Q$-parameters $\Theta _{M}$ are related to the observed VVIX index
through (\ref{linear relationship setup}). We thus use random walk
metropolis method to sample these parameters with the target density as%
\begin{equation*}
\frac{1}{\sqrt{2\pi }\sigma _{P}}\exp \left( -\frac{\sum_{i=1}^{T}\left(
VVIX_{i\Delta }^{2}-A\left( \tau \right) -B\left( \tau \right) \omega
_{i\Delta }\right) ^{2}}{2\sigma _{P}^{2}}\right)
\end{equation*}

For the pricing error parameter $\Theta _{E}=\left\{ \sigma _{P}\right\} $,
conditional on $VVIX_{i\Delta }^{2}$ and $\omega _{i\Delta }$, $\epsilon
_{i\Delta }=VVIX_{i\Delta }^{2}-A\left( \tau \right) -B\left( \tau \right)
\omega _{i\Delta }\sim N\left( 0,\sigma _{P}^{2}\right) $. Assume the prior
for $\sigma _{P}^{2}$ is $\pi _{\sigma _{P}^{2}}\left( \sigma
_{P}^{2}\right) \sim InvGam\left( \alpha _{\sigma _{P}^{2}1},\alpha _{\sigma
_{P}^{2}2}\right) $, we then sample $\sigma _{P}^{2}$ using $InvGam\left(
\alpha _{\sigma _{P}^{2}1}^{\ast },\alpha _{\sigma _{P}^{2}2}^{\ast }\right)
$ with $\alpha _{\sigma _{P}^{2}1}^{\ast }=\alpha _{\sigma _{P}^{2}1}+\frac{%
T-1}{2}$ and $\alpha _{\sigma _{P}^{2}2}^{\ast }=\alpha _{\sigma _{P}^{2}2}+%
\frac{\sum_{i=2}^{T}\left( VVIX_{i\Delta }^{2}-A\left( \tau \right) -B\left(
\tau \right) \omega _{i\Delta }\right) ^{2}}{2}$.

\subsection{Model diagnostics and specification tests}

\subsubsection{Residual analysis}

Given the sampled posterior latent variables (spot volatility, jump times
and jump sizes) and parameters, we can construct several statistics to test
and assess the ability of the model to fit historical data. Recall the
discretization of $Y\left( t\right) $ during the MCMC estimation%
\begin{equation*}
\widetilde{Y}_{i\Delta }=a_{0}+a_{1}Y_{\left( i-1\right) \Delta
}+a_{2}\omega _{\left( i-1\right) \Delta }+\sqrt{\omega _{\left( i-1\right)
\Delta }\Delta }\epsilon _{i\Delta }^{y}
\end{equation*}%
where $\widetilde{Y}_{i\Delta }=Y_{i\Delta }-j_{i\Delta }^{y}n_{i\Delta }$
for $2\leq i\leq T+1$\ and $a_{0}=\kappa _{V}\theta \Delta $, $%
a_{1}=1-\kappa _{V}\Delta $, $a_{2}=-\varsigma _{V}\Delta $. The
representation of $\epsilon _{i\Delta }^{y}$ follows immediately and given by%
\begin{equation}
\epsilon _{i\Delta }^{y}=\frac{\widetilde{Y}_{i\Delta }-a_{0}-a_{1}Y_{\left(
i-1\right) \Delta }-a_{2}\omega _{\left( i-1\right) \Delta }}{\sqrt{\omega
_{\left( i-1\right) \Delta }\Delta }},2\leq i\leq T+1
\label{VIX residual formula}
\end{equation}%
With the estimated variables and parameters at hand we can calculated these
residuals immediately. We will compare the Q-Q plot of the residuals of
different models. If these residuals follow standard normal distribution
approximately, then model performs well for fitting historical VIX index.\
If there exists big discrepancy between the residuals and standard normal
distribution, the corresponding model must have potential for further
improvement. With%
\begin{equation*}
\widetilde{\omega }_{i\Delta }=c_{0}+c_{1}\omega _{\left( i-1\right) \Delta
}+\sigma _{\omega }\sqrt{\omega _{\left( i-1\right) \Delta }\Delta }\epsilon
_{i\Delta }^{\omega }
\end{equation*}%
where $\widetilde{\omega }_{i\Delta }=\omega _{i\Delta }-j_{i\Delta
}^{\omega }n_{i\Delta }$ for $2\leq i\leq T$, and $c_{0}=\alpha _{\omega
}\Delta $, $c_{1}=1-\kappa _{\omega }^{P}\Delta $. We can calculate the
residual for $\omega \left( t\right) $ similarly%
\begin{equation}
\epsilon _{i\Delta }^{\omega }=\frac{\widetilde{\omega }_{i\Delta
}-c_{0}-c_{1}\omega _{\left( i-1\right) \Delta }}{\sigma _{\omega }\sqrt{%
\omega _{\left( i-1\right) \Delta }\Delta }},2\leq i\leq T
\label{volatility residual formula}
\end{equation}%
This residual can be used to compare the various model for $\omega \left(
t\right) $.

The jump times of $Y\left( t\right) $ and $\omega \left( t\right) $ can be
used to test whether the jump intensity is constant or stochastic. If the
posterior sampled jump times are clustered, the constant jump intensity
assumption is rejected.

\subsubsection{$p-$value method}

We also perform simulation study using the posterior parameters to test
different specifications. We first specify some statistics that can reflect
the dynamics of VIX and\ calculate these statistics for the logVIX data.
Then\ for every model, we simulate many trajectories for $Y$ with the same
sample size as the VIX data using the estimated parameters from MCMC
results. With the simulated trajectory, we calculate the sample statistics
and compare them with that obtained from original VIX data. More
specifically, we use the following reference statistics

\begin{itemize}
\item standard deviation

\item skewness

\item kurtosis

\item maximum

\item minimum

\item maxjump: the highest positive changes in the index

\item minjump: the highest negative changes in the index

\item avgmax10: the average over the 10 largest positive changes

\item avgmin10: the average over the 10 largest negative changes

\item various percentiles of daily changes. The percentiles are denoted by $%
percNUM$ where $NUM$ indicates the percentage level.
\end{itemize}

Denote these statistics calculated from logVIX data by $\phi
_{k},k=1,2,\cdots ,10$. Then, for a given model, simulate $N$ trajectories
for $Y$ using the estimated parameters from MCMC results. For the $n$th
simulated trajectory $Y$, $1\leq n\leq N$, calculate the statistics above
which is denoted by $\phi _{k}^{\left( n\right) },k=1,2,\cdots ,10$. For
every $k$, $1\leq k\leq 10$, calculate%
\begin{equation}
p_{k}=\frac{\sum_{n=1}^{N}1_{\left\{ \phi _{k}^{\left( n\right) }>\phi
_{k}\right\} }}{N}  \label{p-value}
\end{equation}%
where $1_{A}$ is the indicator function. Too high or too low $p_{k},1\leq
k\leq 10$ indicates that the given model may distort from the genuine form.
For more details about this method, we refer to \cite{Gelman, Meng and Stein (1996)} and \cite{Kaeck and Alexander (2013)}.

\section{Data}

\label{section:data description}

In 1993, CBOE introduced VIX index and it serves as a benchmark for the
volatility of the market. In September 22, 2003, the CBOE revised the
calculation method of VIX which utilized a wider range of S\&P500 options
and back-calculated the new VIX to 1990.\ The well-known generalized formula
now for calculating VIX is%
\begin{equation*}
VIX^{2}\left( t,T\right) =\frac{2}{T-t}\sum_{i}\frac{\bigtriangleup K_{i}}{%
K_{i}^{2}}e^{r_{t}\left( T-t\right) }Q\left( K_{i}\right) -\frac{1}{T-t}%
\left[ \frac{F_{t}}{K_{0}}-1\right] ^{2}
\end{equation*}%
which utilizes a strip of OTM S\&P500 options prices $Q\left( K_{i}\right) $
and $F_{t}$ is the forward S\&P500 index level derived from S\&P500 options.

\begin{figure}[htb]
\par
\begin{center}
{\ \centering
\includegraphics[scale=0.5,bb=0 0 800 780]{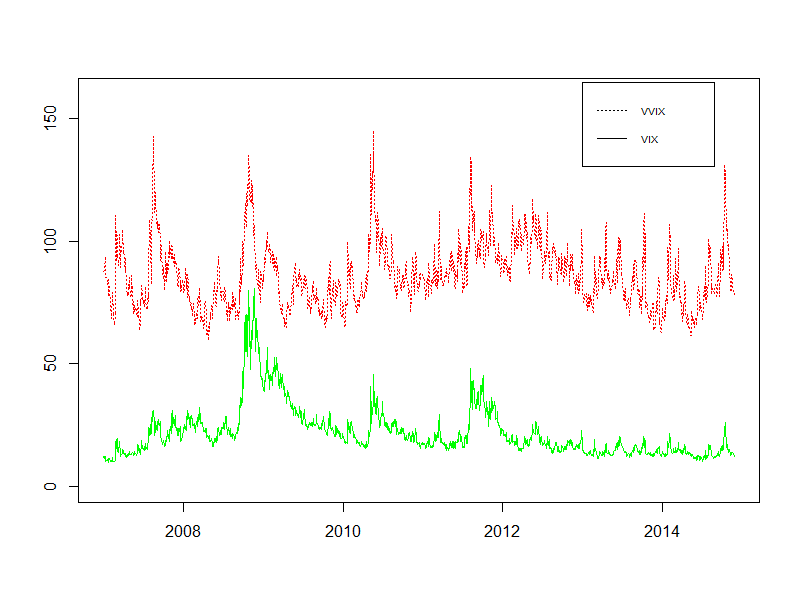} }
\end{center}
\par
{\scriptsize
\flushleft{This figure shows the time series of VIX and VVIX
index from January 3, 2007 to November 26, 2014. Both of them are mean-reverting and VVIX is at a significant higher level than VIX in terms of the range of values.}%
}
\caption{VIX and VVIX index}
\label{VIX-VVIX}
\end{figure}

In March 14, 2012, CBOE released a new volatility of volatility index called
VVIX. VVIX is a measure of volatility of volatility which represents the
expected volatility of the 30-day forward price of the CBOE volatility
index. The calculation method of VVIX is similar to VIX, it is calculated
from the price of a strip of at- and out- of the money VIX options, i.e.%
\begin{equation*}
VVIX^{2}\left( t,T\right) =\frac{2}{T-t}\sum_{i}\frac{\bigtriangleup K_{i}}{%
K_{i}^{2}}e^{r_{t}\left( T-t\right) }O\left( K_{i}\right) -\frac{1}{T-t}%
\left[ \frac{F_{t}}{K_{0}}-1\right] ^{2}
\end{equation*}%
where $O\left( K_{i}\right) $ is the midpoint of the bid-ask spread for VIX
option with strike $K_{i}$ and $F_{t}$ is the forward VIX index level
derived from VIX option prices. $K_{0}$ is the first strike below the
forward index level $F_{t}$. Using this method, CBOE has also calculated the
VVIX index before the release data up to the start of 2007. We plot the
historical time series of VIX and VVIX from Jan 2007 to Nov 2014 in Figure %
\ref{VIX-VVIX}. From the picture we can see that the range of VVIX is at a
significantly higher level than that of the VIX. Like VIX, VVIX also mean
reverts to its historical mean value which is nearly 80. Furthermore, they
share some of their peak values, especially during the 2008 financial
crisis. Compared with VIX, VVIX is more volatile and when VIX is high, the
range of variation of VVIX widens. The statistics of VIX and VVIX from Jan
2007 to Nov 2014 are summarized in Table \ref{Summary Statistics}

\section{Empirical results}

\label{section:empirical results}

In this section we discuss the estimation result for VIX dynamics among
different models. The parameter estimation for four models are summarized in
Table \ref{parameter estimation result} and the simulation results are
showed in Table \ref{Simulation results}. For all of the candidate models,
the estimates of $\rho $ are positive and around 0.52 which is close to the
result in \cite{Kaeck and Alexander (2013)} ($\rho =0.659$ for SVJ model in their
paper). As the parameter $\varsigma _{V}$ enters into the drift of the VIX,
the estimation for $\theta $ is relatively low compared to the mean value of
logVIX market data during the same period. $\kappa _{\omega }^{P}$ is
significantly larger than $\kappa _{V}$ and this reflects the fact that the
volatility of VIX or VVIX is more volatile than VIX itself.

\begin{figure}[]
\par
\begin{center}
{\ \centering
\includegraphics[scale=0.5,bb=0 0 900 680]{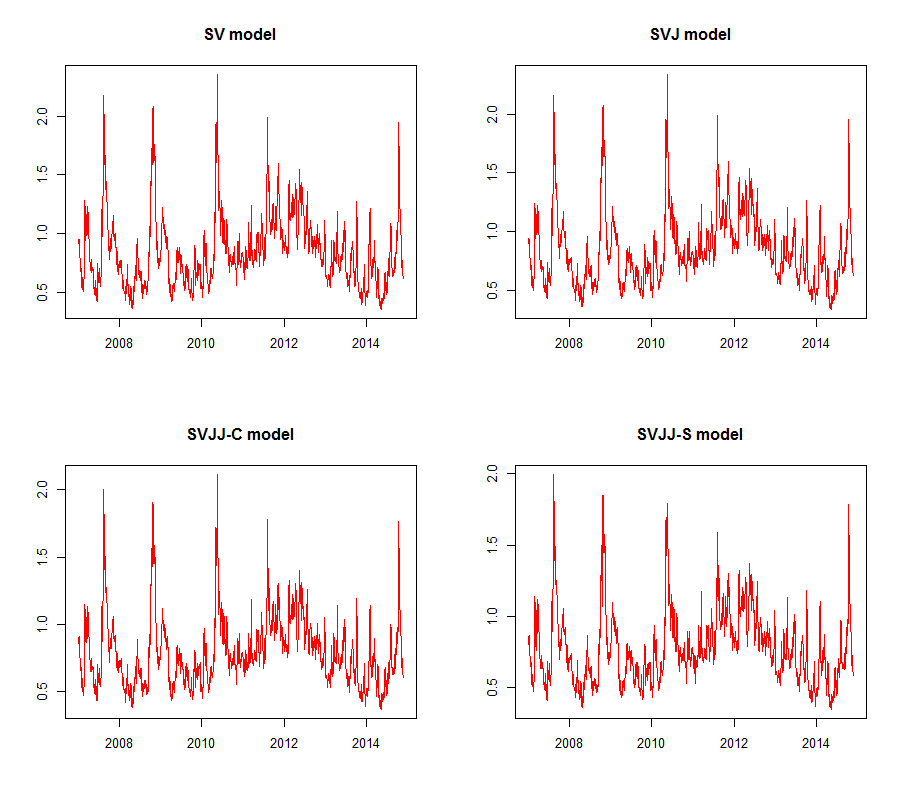} }
\end{center}
\par
{\scriptsize
\flushleft{The figures show the estimated paths of posterior
volatility $\omega \left( t\right)$ for four models. All of them are highly
correlated with VVIX index. The level of the volatility in SV and SVJ models
is slightly higher than that in SVJJ-C and SVJJ-S models. "SV" denotes
diffusion model with no jumps. "SVJ" introduces jumps in VIX in the SV model
with constant jump intensity, "SVJJ-C" adds double jumps in VIX and its
volatility in the SV model with constant jump intensity. "SVJJ-S" assumes
the jump intensity to be stochastic in the SVJJ-C model.}%
}
\caption{Posterior volatility of VIX for each of the models}
\label{Posterior volatility of VIX for each of the models}
\end{figure}

Figure \ref{Posterior volatility of VIX for each of the models} gives the
estimated volatility process of VIX for four models. As we use VVIX index as
a proxy for the volatility, this four processes present similar forms. The
correlation between the estimated spot volatility and VVIX of the four
models are 0.9781, 0.9782, 0.9822 and 0.9807 respectively. The average
posterior volatility for SV and SVJ models is slightly higher compared to
the other two models. This can be explained that the addition of jumps in
volatility reduces the demand on the volatility process.

\begin{figure}[]
\par
\begin{center}
{\ \centering
\includegraphics[scale=0.5,bb=0 0 900 680]{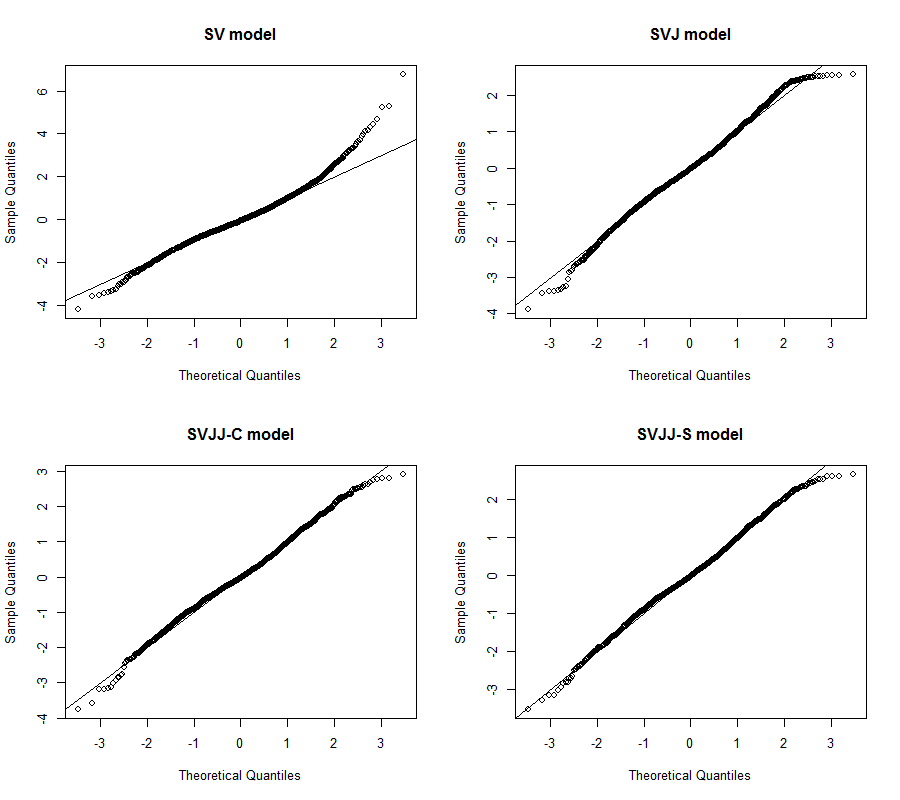} }
\end{center}
\par
{\scriptsize
\flushleft{The figures show the Q-Q plot of the residuals
calculated from each of the models using (\ref{VIX residual formula}) with the estimated parameters as
input. SVJJ-C and SVJJ-S models perform relatively better than SV and SVJ
models. "SV" denotes diffusion model with no jumps. "SVJ" introduces jumps
in VIX in the SV model with constant jump intensity, "SVJJ-C" adds double
jumps in VIX and its volatility in the SV model with constant jump
intensity. "SVJJ-S" assumes the jump intensity to be stochastic in the
SVJJ-C model.}%
}
\caption{Q-Q plot of the residuals}
\label{Q-Q plot of the residuals}
\end{figure}

\begin{figure}[]
\par
\begin{center}
{\ \centering
\includegraphics[scale=0.5,bb=0 0 900 680]{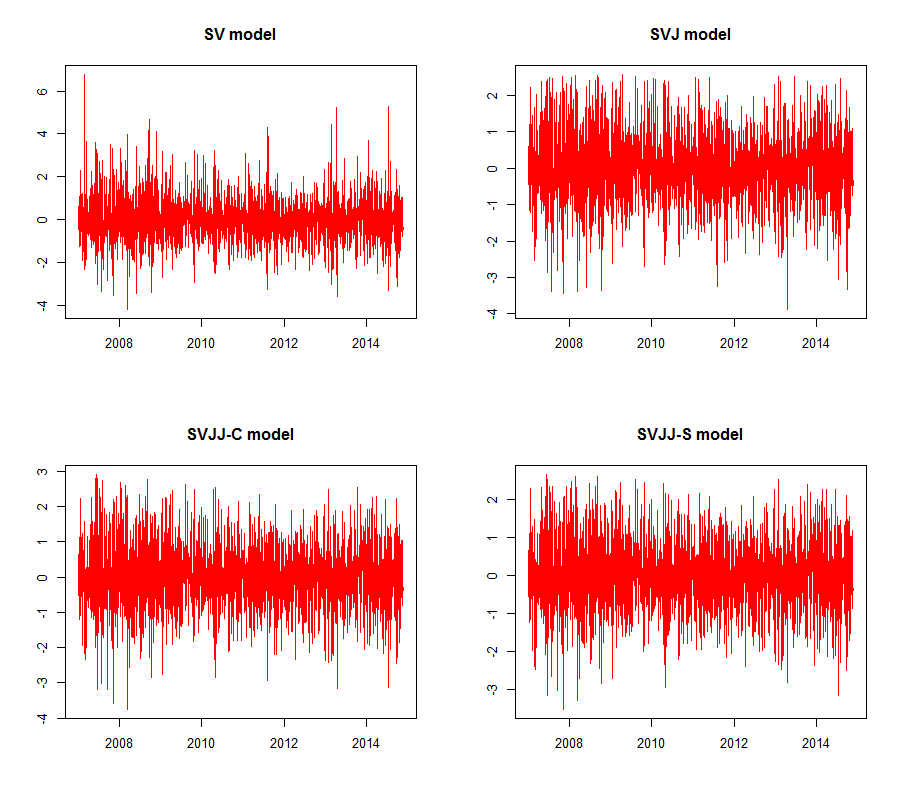} }
\end{center}
\par
{\scriptsize
\flushleft{The figures show the time series of standard innovations or residuals of VIX
calculated from the estimated parameters using (\ref{volatility residual formula}). "SV" denotes diffusion model with no jumps. "SVJ" introduces jumps in VIX in the SV model with constant jump intensity, "SVJJ-C" adds double jumps in VIX and its volatility in the SV model with constant jump intensity. "SVJJ-S" assumes the jump intensity to be stochastic in the SVJJ-C model.}%
}
\caption{VIX Residuals}
\label{VIX Residuals}
\end{figure}

Figure \ref{Q-Q plot of the residuals} shows the Q-Q plot of the residuals
of VIX calculated by (\ref{VIX residual formula}) for all of the four models
and Figure~\ref{VIX Residuals} plot the time series form. From the upper
left panel in Figure~\ref{Q-Q plot of the residuals} we find the SV model is
misspecified for it requires very large shocks to Brownian motion. This can
also be seen from the upper left panel in Figure~\ref{VIX Residuals}.
Compared to the other three models, the range of the residuals for SV model
is significantly larger and there exists many large innovations.

\begin{figure}[]
\par
\begin{center}
{\ \centering
\includegraphics[scale=0.5,bb=0 0 800 780]{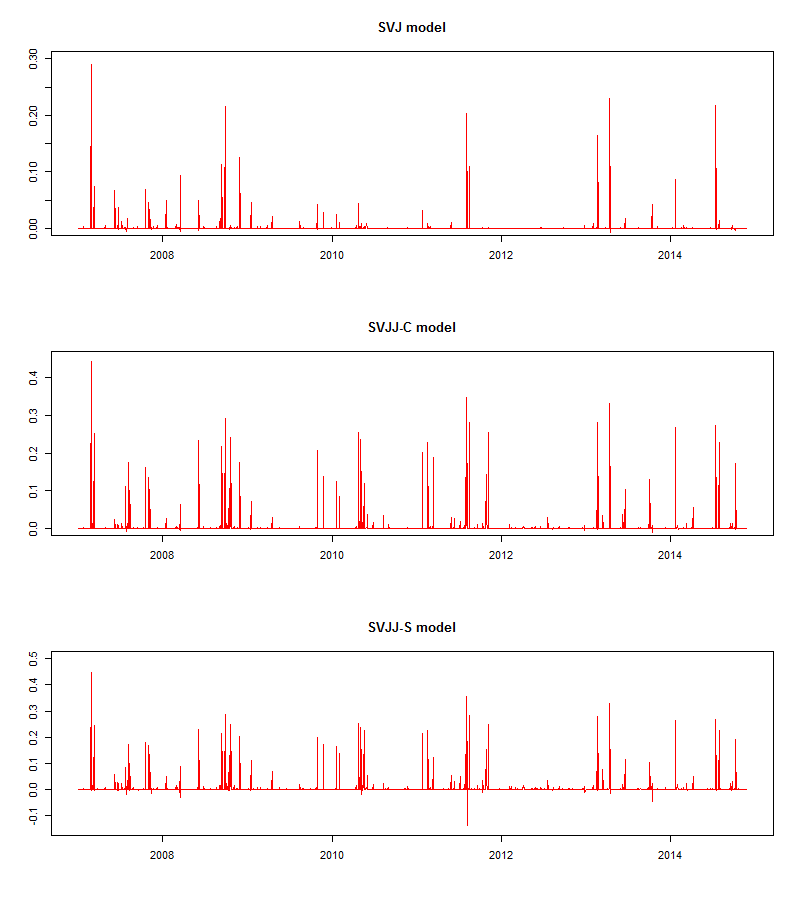} }
\end{center}
\par
{\scriptsize
\flushleft{The figures show the time series of average jump
sizes in VIX. "SV" denotes diffusion model with no jumps. "SVJ" introduces
jumps in VIX in the SV model with constant jump intensity, "SVJJ-C" adds
double jumps in VIX and its volatility in the SV model with constant jump
intensity. "SVJJ-S" assumes the jump intensity to be stochastic in the
SVJJ-C model.}%
}
\caption{Posterior Mean of Jumps in VIX}
\label{Posterior Mean of Jumps in VIX}
\end{figure}

\begin{figure}[]
\par
\begin{center}
{\ \centering
\includegraphics[scale=0.5,bb=0 0 900 680]{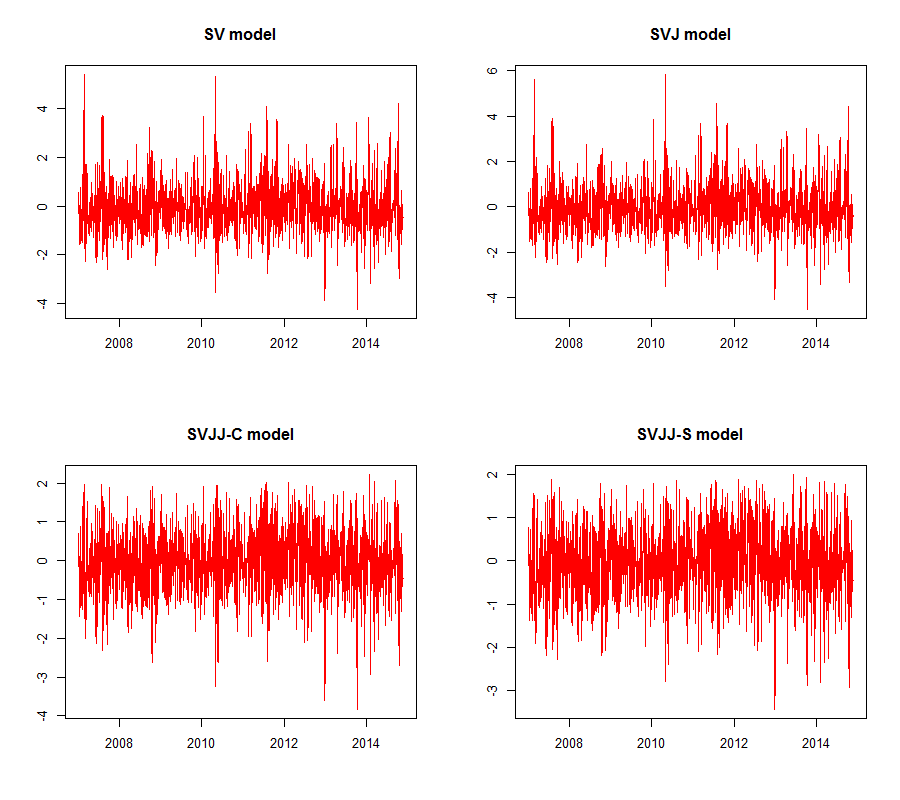} }
\end{center}
\par
{\scriptsize
\flushleft{The figures show the time series of standard innovations or residuals of
volatility of VIX calculated from the estimated parameters using . "SV" denotes diffusion model with no jumps. "SVJ" introduces jumps in VIX in the SV model with constant jump intensity, "SVJJ-C" adds double jumps in VIX and its volatility in the SV model with constant jump intensity. "SVJJ-S" assumes the jump intensity to be stochastic in the SVJJ-C model.}%
}
\caption{Volatility Residuals}
\label{Volatility Residuals}
\end{figure}

From the two upper panels in Figure~\ref{Q-Q plot of the residuals} we can
see the tail of the residuals become slightly thin so SVJ model improves SV
model better. Much of the big Brownian shocks can be absorbed into the jump
part. The estimated jump size in VIX is reported in the upper left panel in
Figure~\ref{Posterior Mean of Jumps in VIX}. However, from the simulation
results in Table \ref{Simulation results}, we find that for both SV and SVJ
model, there are one or more statistics whose $p$-value are out of the $%
\left[ 0.05,0.95\right] $ bound. In contrast, in \cite{Kaeck and Alexander (2013)},
they also test the SV and SVJ model (with normal jump) and demonstrate that
all of the $p$-value are within the $\left[ 0.05,0.95\right] $ bound. This
shows that the addition of VVIX as the poxy for volatility of VIX help
detect the further space of improvement for the stochastic volatility of
volatility model for VIX. We also turn to Figure~\ref{Volatility Residuals}
calculated using (\ref{volatility residual formula}) which compares the
residuals of volatility processes among four models. The residuals of SV and
SVJ models are evidently larger than SVJJ-C and SVJJ-S model.

\begin{figure}[]
\par
\begin{center}
{\ \centering
\includegraphics[scale=0.5,bb=0 0 800 780]{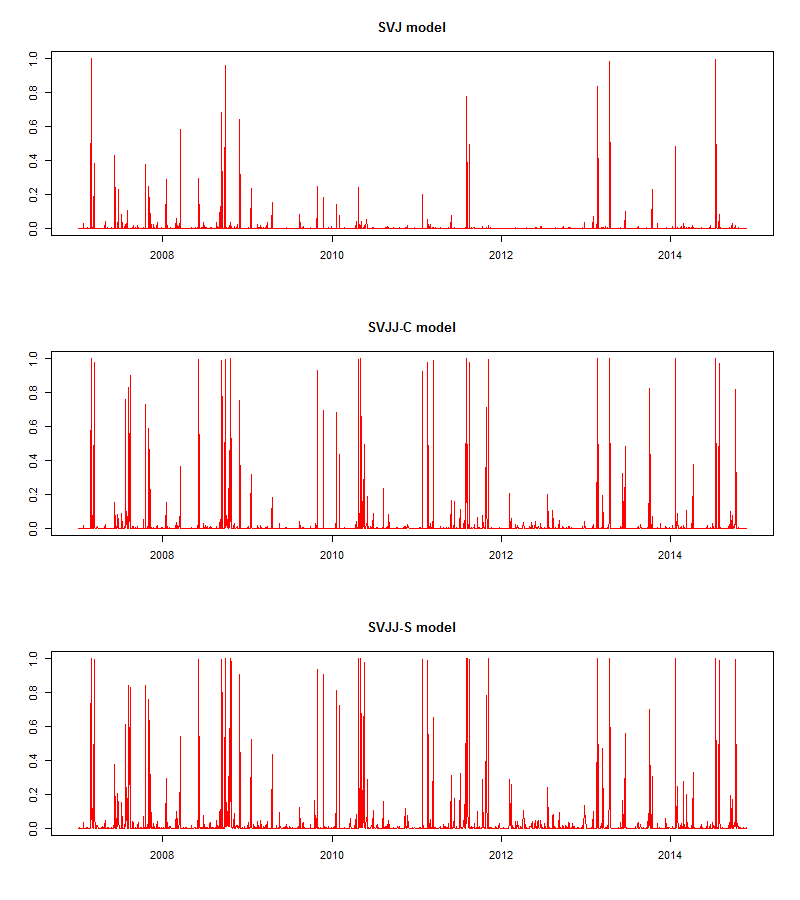} }
\end{center}
\par
{\scriptsize
\flushleft{The figures show the estimated jump probability of
SVJ, SVJJ-C and SVJJ-S models. "SVJ" introduces jumps in VIX with constant
jump intensity and models the volatility using square root diffusion model,
"SVJJ-C" introduces double jumps in VIX and its volatility with constant
jump intensity. "SVJJ-S" assumes the jump intensity to be stochastic in the
SVJJ-C model.}%
}
\caption{Estimated jump times for SVJ, SVJJ-C and SVJJ-S models}
\label{Estimated jump times}
\end{figure}

Next we come to the SVJJ-C and SVJJ-S models. As mentioned above, the
residuals of volatility processes for this two models performs better than
SV and SVJ models. This shows the impact of jump on the volatility. Figure %
\ref{Estimated jump times} describe the estimated daily jump probability for
SVJ, SVJJ-C and SVJJ-S models. Evidently, with the jump in volatility added,
the jump occurs a bit more frequently. We recall that for SVJ model in which
no jump happens in volatility, the jump time are determined mainly by the
information of the VIX index. While for SVJJ-C and SVJJ-S models, we sample
the jump time using both information or signal from VIX and volatility
(VVIX). As we assume that the jumps of VIX and its volatility factor are
determined by the same Poisson process, a big jump in volatility may raise
the jump probability. This means that not only does the volatility jump, but
furthermore it jumps more heavily than VIX. Return to Figure~\ref{Q-Q plot
of the residuals}, the bottom panels for SVJJ-C and SVJJ-S performs better
than the upper ones, this show that the jump in volatility can also have
impact on the dynamics of VIX. The influence channel can be through moments
of high order or extreme values which can be seen from Table \ref{Simulation
results}.

Unlike transient Brownian motion shocks, the influence of jump in volatility
is more persistent. After positive or negative jump, the volatility enters a
new regime. As the diffusion part of VIX, its effect will last for a period.
A simple empirical method for judging the existence of jump in volatility in
some day is to compare the fluctuation of a period of VIX data before and
after that day.\ For example, on Feb 27, 2007, the VIX jumped from 11.15 to
18.31. Before this day for a long period, the VIX looked very tranquil with
very negligible variation and stay around 11. However, after this turning
point, the VIX became more volatile and large up or down occurred more
frequently, ranging from 12.19 to 19.63 during the next 20 days. In fact, on
Feb 27 the VVIX index also jumped from 70.33 to 110.42. If we calculate the
average of VVIX index for 20 days before and after this day, the results are
72.54 and 96.55 respectively. This indicate that the volatility had changed
from the original state to a new and higher regime and thus made VIX index
more active. This effect cannot be achieved by just a single Brownian shock
on volatility and should be caused by jump. Thus the SVJ model is
misspecified also from empirical observation.

\begin{figure}[]
\par
\begin{center}
{\ \centering
\includegraphics[scale=0.5,bb=0 0 800 680]{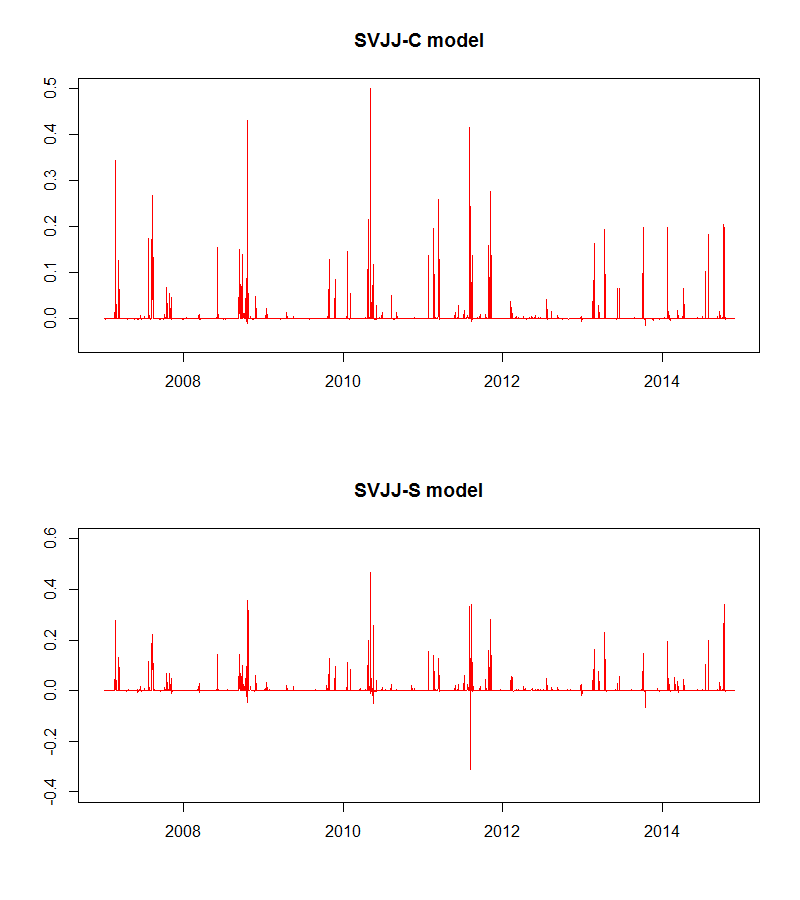} }
\end{center}
\par
{\scriptsize
\flushleft{The figures show the time series of average jump
sizes in the volatility of VIX. "SVJJ-C" introduces double jumps in VIX and
its volatility with constant jump intensity. "SVJJ-S" assumes the jump
intensity to be stochastic in the SVJJ-C model.}%
}
\caption{Posterior Mean of Jumps in volatility of VIX}
\label{Posterior Mean of Jumps in volatility of VIX}
\end{figure}

Figure~\ref{Posterior Mean of Jumps in volatility of VIX} describes the jump
of volatility for SVJJ-C and SVJJ-S models respectively. The jump size are
almost positive with only a big negative jump in SVJJ-S models. With the
mean-reverting property, the volatility reduces to its mean level through
negative Brownian innovations after a big positive jump. This also indicates
that the impact of positive jump can be persistent and significant.

From Figure~\ref{Estimated jump times}, we observe that for SVJJ-C model,
the jump times are clustered. This is extremely unlikely under the constant
jump intensity assumption. We can also see from Table \ref{parameter
estimation result} that the estimation of $\lambda _{1}$ in SVJJ-S model is
significant above zero. These facts indicate that the SVJJ-S model is
superior to SVJJ-C model and depict the dynamics of VIX more accurately.
When the stochastic volatility $\omega \left( t\right) $ enters into a
relatively high regime, more jumps happen and affect the dynamics of VIX.

\section{Conclusion}

\label{section:conclusion}

This paper discusses the model specification for stochastic volatility
models of VIX from information of VVIX. We construct a volatility proxy of
VIX using VVIX index as the benchmark and study its role for improving the
model assumption of VIX from empirical observations. Based on the joint
behavior of VIX and VVIX we propose a double-jump stochastic volatility
model for VIX. We use MCMC method to estimate and compare different nested
models using daily data of VIX and VVIX. Based on this, we point out that
the jumps in VIX and volatility are essential and statistically significant
and analyze the impact of the jumps on VIX dynamics. We show the jump
intensity is stochastic and state dependent. The use of VVIX brings the
estimation of some $Q$-parameters. Compared with richer dataset composed of
VIX futures and options, the accuracy of these parameters have potential for
further improvement. The corresponding risk premia could be further
specified. This will be left for future work.



\section*{Appendix: MCMC algorithm for inference}

The MCMC sampling methods for parameters under physical measure $P$, jump
times and jump sizes\ provided here are standard. Our sampling algorithms
here borrow from \cite{Johannes and Polson (2003)}, \cite{Kaeck and Alexander (2010)} and \cite{Amengual and Xiu (2012)}. To set up, the jump-adjusted discretization of the
processes under $P$ are stated as follows:%
\begin{eqnarray*}
\widetilde{Y}_{i\Delta } &=&a_{0}+a_{1}Y_{\left( i-1\right) \Delta
}+a_{2}\omega _{\left( i-1\right) \Delta }+\sqrt{\omega _{\left( i-1\right)
\Delta }\Delta }\epsilon _{i\Delta }^{y} \\
\widetilde{\omega }_{i\Delta } &=&c_{0}+c_{1}\omega _{\left( i-1\right)
\Delta }+\sigma _{\omega }\sqrt{\omega _{\left( i-1\right) \Delta }\Delta }%
\epsilon _{i\Delta }^{\omega }
\end{eqnarray*}%
where $a_{0}=\kappa _{V}\theta \Delta $, $a_{1}=1-\kappa _{V}\Delta $, $%
a_{2}=-\varsigma _{V}\Delta $, $c_{0}=\alpha _{\omega }\Delta $, $%
c_{1}=1-\kappa _{\omega }^{P}\Delta $ and $\widetilde{Y}_{i\Delta
}=Y_{i\Delta }-j_{i\Delta }^{y}n_{i\Delta }$, $\widetilde{\omega }_{i\Delta
}=\omega _{i\Delta }-j_{i\Delta }^{\omega }n_{i\Delta }$. Our aim is to
estimate $P$-parameters%
\begin{equation*}
\Theta _{P}=\left\{ \kappa _{V},\varsigma _{V},\theta ,\kappa _{\omega
}^{P},\mu _{y}^{JP},\mu _{\omega }^{JP},\sigma _{\omega }^{J},\rho ,\sigma
_{\omega }\right\}
\end{equation*}%
and latent variables $n_{i\Delta },j_{i\Delta }^{y}$ and $j_{i\Delta
}^{\omega },$ $2\leq i\leq T+1$.

\subsection*{A. Sampling latent variables}

\begin{itemize}
\item {Sample Jump times}

\bigskip\ \ For $n_{i\Delta },i=2,3,\cdots ,T,$%
\begin{eqnarray*}
&&p\left( n_{i\Delta }=1|X,\Theta _{P},Y\right) \\
&=&\frac{p\left( Y_{i\Delta },\omega _{i\Delta }|Y_{\left( i-1\right) \Delta
},\omega _{\left( i-1\right) \Delta },n_{i\Delta }=1,j_{\left( i-1\right)
\Delta }^{y},j_{\left( i-1\right) \Delta }^{\omega },\Theta _{P}\right)
\cdot p\left( n_{i\Delta }=1|\omega _{\left( i-1\right) \Delta },Y_{\left(
i-1\right) \Delta }\right) }{\sum\limits_{s=0}^{1}p\left( Y_{i\Delta
},\omega _{i\Delta }|Y_{\left( i-1\right) \Delta },\omega _{\left(
i-1\right) \Delta },n_{i\Delta }=s,j_{\left( i-1\right) \Delta
}^{y},j_{\left( i-1\right) \Delta }^{\omega },\Theta _{P}\right) \cdot
p\left( n_{i\Delta }=s|\omega _{\left( i-1\right) \Delta },Y_{\left(
i-1\right) \Delta }\right) }
\end{eqnarray*}%
where $p\left( Y_{i\Delta },\omega _{i\Delta }|Y_{\left( i-1\right) \Delta
},\omega _{\left( i-1\right) \Delta },n_{i\Delta }=s,j_{\left( i-1\right)
\Delta }^{y},j_{\left( i-1\right) \Delta }^{\omega },\Theta _{P}\right) $ is
a bivariate normal distribution with mean%
\begin{equation*}
\left[
\begin{array}{c}
a_{0}+a_{1}Y_{\left( i-1\right) \Delta }+a_{2}\omega _{\left( i-1\right)
\Delta }+s\cdot j_{i\Delta }^{y} \\
c_{0}+c_{1}\omega _{\left( i-1\right) \Delta }+s\cdot j_{i\Delta }^{\omega }%
\end{array}%
\right]
\end{equation*}%
and covariance matrix%
\begin{equation*}
\omega _{\left( i-1\right) \Delta }\Delta \left[
\begin{array}{cc}
1 & \rho \sigma _{\omega } \\
\rho \sigma _{\omega } & \sigma _{\omega }^{2}%
\end{array}%
\right]
\end{equation*}%
and $p\left( n_{i\Delta }=1|\omega _{\left( i-1\right) \Delta },Y_{\left(
i-1\right) \Delta }\right) =\left( \lambda _{0}+\lambda _{1}\omega _{\left(
i-1\right) \Delta }\right) \Delta $.

\ For $n_{(T+1)\Delta }$,%
\begin{eqnarray*}
&&p\left( n_{(T+1)\Delta }=1|X,\Theta _{P},Y\right) \\
&\varpropto &\frac{p\left( Y_{(T+1)\Delta }|Y_{T\Delta },\omega _{T\Delta
},n_{(T+1)\Delta }=1,j_{T\Delta }^{y},j_{T\Delta }^{\omega },\Theta
_{P}\right) \cdot p\left( n_{(T+1)\Delta }=1|Y_{T\Delta }\right) }{%
\sum\limits_{s=0}^{1}p\left( Y_{(T+1)\Delta }|Y_{T\Delta },\omega _{T\Delta
},n_{(T+1)\Delta }=s,j_{T\Delta }^{y},j_{T\Delta }^{\omega },\Theta
_{P}\right) \cdot p\left( n_{(T+1)\Delta }=s|Y_{T\Delta }\right) }
\end{eqnarray*}%
where $p\left( Y_{(T+1)\Delta }|Y_{T\Delta },\omega _{T\Delta
},n_{(T+1)\Delta }=s,j_{T\Delta }^{y},j_{T\Delta }^{\omega },\Theta
_{P}\right) $ is a univariate normal distribution with mean $%
a_{0}+a_{1}Y_{T\Delta }+a_{2}\omega _{T\Delta }+s\cdot j_{(T+1)\Delta }^{y}$
and variance $\omega _{T\Delta }\Delta $ and $p\left( n_{(T+1)\Delta
}=1|Y_{T\Delta }\right) =\left( \lambda _{0}+\lambda _{1}\omega _{T\Delta
}\right) \Delta $. We can thus sample $n_{i\Delta }$ for $i=2,3,\cdots ,T+1$.

\item {Sample Jump Sizes}

\ When $n_{i\Delta }=1$, we then sample $j_{i\Delta }^{\omega }$ using $%
N\left( \frac{B}{A},\frac{1}{A}\right) $ where%
\begin{eqnarray*}
A &=&\frac{1}{\sigma _{\omega }^{2}\omega _{\left( i-1\right) \Delta }\Delta
}+\frac{1}{\left( \sigma _{\omega }^{J}\right) ^{2}} \\
B &=&\frac{\omega _{i\Delta }-c_{0}-c_{1}\omega _{\left( i-1\right) \Delta }%
}{\sigma _{\omega }^{2}\omega _{\left( i-1\right) \Delta }\Delta }+\frac{\mu
_{\omega }^{JP}}{\left( \sigma _{\omega }^{J}\right) ^{2}}
\end{eqnarray*}

and sample $j_{i\Delta }^{y}$ using $N\left( \frac{B}{A},\frac{1}{A}\right) $
where%
\begin{eqnarray*}
A &=&\frac{1}{\omega _{\left( i-1\right) \Delta }\left( 1-\rho ^{2}\right)
\Delta }+\frac{1}{\left( \sigma _{y}^{J}\right) ^{2}} \\
B &=&\frac{Y_{i\Delta }-a_{0}-a_{1}Y_{\left( i-1\right) \Delta }-a_{2}\omega
_{\left( i-1\right) \Delta }-\frac{\rho }{\sigma _{\omega }}\left( \omega
_{i\Delta }-c_{0}-c_{1}\omega _{\left( i-1\right) \Delta }-j_{i\Delta
}^{\omega }\right) }{\omega _{\left( i-1\right) \Delta }\left( 1-\rho
^{2}\right) \Delta }+\frac{\mu _{y}^{JP}}{\left( \sigma _{y}^{J}\right) ^{2}}
\end{eqnarray*}

\ When $n_{i\Delta }=0$, the posterior distribution of $j_{i\Delta }^{y}$
and $j_{i\Delta }^{\omega }$ are same with the prior distribution, i.e.,%
\begin{eqnarray*}
&&J_{Y}^{P}\sim N\left( \mu _{y}^{JP},\left( \sigma _{y}^{J}\right)
^{2}\right) \\
&&J_{\omega }^{P}\text{ }\symbol{126}\text{ }N\left( \mu _{\omega
}^{JP},\left( \sigma _{\omega }^{J}\right) ^{2}\right)
\end{eqnarray*}%
where the parameters are updated simultaneously. \ \

\ We thus sample $j_{i\Delta }^{y}$ and $j_{i\Delta }^{\omega }$ for $%
i=2,3,\cdots ,T+1$.
\end{itemize}

\subsection*{B. Sampling parameters $\Theta _{P}$}

The parameter set under $P$ is denoted by%
\begin{equation*}
\Theta _{P}=\left\{ \kappa _{V},\varsigma _{V},\theta ,\kappa _{\omega
}^{P},\mu _{y}^{JP},\mu _{\omega }^{JP},\sigma _{\omega }^{J},\rho ,\sigma
_{\omega }\right\}
\end{equation*}

\begin{itemize}
\item {Sampling $\theta $}

Assume the prior for $\theta $ is $N\left( \mu _{\theta },\sigma _{\theta
}^{2}\right) $, we sample the posterior using $\theta \sim N(B/A,1/A)$ with%
\begin{eqnarray*}
A &=&\frac{1}{\sigma _{\theta }^{2}}+\sum_{i=2}^{T}\frac{\kappa
_{V}^{2}\Delta }{\left( 1-\rho ^{2}\right) \omega _{\left( i-1\right) \Delta
}} \\
B &=&\frac{\mu _{\theta }}{\sigma _{\theta }^{2}}+\kappa _{V}\sum_{i=2}^{T}%
\frac{\widetilde{Y}_{i\Delta }-Y_{\left( i-1\right) \Delta }+\kappa
_{V}Y_{\left( i-1\right) \Delta }\Delta +\varsigma _{V}\omega _{\left(
i-1\right) \Delta }\Delta -\rho D_{i\Delta }\sqrt{\omega _{\left( i-1\right)
\Delta }\Delta }}{\left( 1-\rho ^{2}\right) \omega _{\left( i-1\right)
\Delta }}
\end{eqnarray*}

\item {Sampling $\kappa _{V}$}

Assume the prior for {$\kappa _{V}$} is $N\left( \mu _{{\kappa _{V}}},\sigma
_{{\kappa _{V}}}^{2}\right) $, we sample the posterior using {$\kappa _{V}$ }%
$\sim N(B/A,1/A)$ with%
\begin{eqnarray*}
A &=&\frac{1}{\sigma _{\kappa _{V}}^{2}}+\sum_{i=2}^{T}\frac{\left( \theta
-Y_{\left( i-1\right) \Delta }\right) ^{2}\Delta }{\left( 1-\rho ^{2}\right)
\omega _{\left( i-1\right) \Delta }} \\
B &=&\frac{\mu _{\kappa _{V}}}{\sigma _{\kappa _{V}}^{2}}+\sum_{i=2}^{T}%
\frac{\left( \widetilde{Y}_{i\Delta }-Y_{\left( i-1\right) \Delta
}+\varsigma _{V}\omega _{\left( i-1\right) \Delta }\Delta -\rho D_{i\Delta }%
\sqrt{\omega _{\left( i-1\right) \Delta }\Delta }\right) \left( \theta
-Y_{\left( i-1\right) \Delta }\right) }{\left( 1-\rho ^{2}\right) \omega
_{\left( i-1\right) \Delta }}
\end{eqnarray*}

\item {Sampling $\varsigma _{V}$}

Assume the prior for {$\varsigma _{V}$} is $N\left( \mu _{{\varsigma _{V}}%
},\sigma _{{\varsigma _{V}}}^{2}\right) $, we sample the posterior using {$%
\varsigma _{V}$}$\sim N(B/A,1/A)$ with%
\begin{eqnarray*}
A &=&\frac{1}{\sigma _{\varsigma _{V}}^{2}}+\sum_{i=2}^{T}\frac{\omega
_{\left( i-1\right) \Delta }\Delta }{1-\rho ^{2}} \\
B &=&\frac{\mu _{\kappa _{V}}}{\sigma _{\varsigma _{V}}^{2}}-\sum_{i=2}^{T}%
\frac{\left( \widetilde{Y}_{i\Delta }-Y_{\left( i-1\right) \Delta }-\kappa
_{V}\left( \theta -Y_{\left( i-1\right) \Delta }\right) \Delta -\rho
D_{i\Delta }\sqrt{\omega _{\left( i-1\right) \Delta }\Delta }\right) }{%
1-\rho ^{2}}
\end{eqnarray*}

\item {Sampling $\kappa _{\omega }^{P}$}

Assume the prior for $\kappa _{\omega }^{P}$ is $N\left( \mu _{\kappa
_{\omega }^{P}},\sigma _{\kappa _{\omega }^{P}}^{2}\right) $, we sample the
posterior using $\kappa _{\omega }^{P}\sim N(B/A,1/A)$ with%
\begin{eqnarray*}
A &=&\frac{1}{\sigma _{\kappa _{\omega }^{P}}^{2}}+\sum_{i=1}^{T}\frac{%
\omega _{\left( i-1\right) \Delta }\Delta }{\left( 1-\rho ^{2}\right) \sigma
_{\omega }^{2}} \\
B &=&\frac{\mu _{\kappa _{\omega }^{P}}}{\sigma _{\kappa _{\omega }^{P}}^{2}}%
-\sum_{i=1}^{T}\frac{\widetilde{\omega }_{i\Delta }-\omega _{\left(
i-1\right) \Delta }-\alpha _{\omega }\Delta -\sigma _{\omega }\rho
C_{i\Delta }\sqrt{\omega _{\left( i-1\right) \Delta }\Delta }}{\left( 1-\rho
^{2}\right) \sigma _{\omega }^{2}}
\end{eqnarray*}

\item {Sampling $\mu _{y}^{JP},\mu _{\omega }^{JP},\left( \sigma _{\omega
}^{J}\right) ^{2}$}

Assume the prior for these parameters are: $\mu _{y}^{JP}\sim N\left( \mu
_{\mu _{y}^{JP}},\sigma _{\mu _{y}^{JP}}^{2}\right) $, $\mu _{\omega
}^{JP}\sim N\left( \mu _{\mu _{\omega }^{JP}},\sigma _{\mu _{\omega
}^{JP}}^{2}\right) $, $\left( \sigma _{\omega }^{J}\right) ^{2}\sim
InvGam\left( \alpha _{\left( \sigma _{\omega }^{J}\right) ^{2}1}^{\ast
},\alpha _{\left( \sigma _{\omega }^{J}\right) ^{2}2}^{\ast }\right) $

Then we sample the posterior using%
\begin{eqnarray*}
\mu _{\omega }^{JP} &\sim &N\left( \frac{\left( \sigma _{\omega }^{J}\right)
^{2}\mu _{\mu _{\omega }^{JP}}+\sigma _{\mu _{\omega
}^{JP}}^{2}\sum_{i=2}^{T}j_{i\Delta }^{\omega }}{\left( \sigma _{\omega
}^{J}\right) ^{2}+T\sigma _{\mu _{\omega }^{JP}}^{2}},\left( \frac{T}{\left(
\sigma _{\omega }^{J}\right) ^{2}}+\frac{1}{\sigma _{\mu _{\omega }^{JP}}^{2}%
}\right) ^{-0.5}\right) \\
\mu _{y}^{JP} &\sim &N\left( \frac{\left( \sigma _{y}^{J}\right) ^{2}\mu
_{\mu _{y}^{JP}}+\sigma _{\mu _{y}^{JP}}^{2}\sum_{i=2}^{T}j_{i\Delta }^{y}}{%
\left( \sigma _{y}^{J}\right) ^{2}+T\sigma _{\mu _{y}^{JP}}^{2}},\left(
\frac{T}{\left( \sigma _{y}^{J}\right) ^{2}}+\frac{1}{\sigma _{\mu
_{y}^{JP}}^{2}}\right) ^{-0.5}\right) \\
\left( \sigma _{\omega }^{J}\right) ^{2} &\sim &InvGam\left( \alpha _{\left(
\sigma _{\omega }^{J}\right) ^{2}1}^{\ast }+\frac{T}{2},\alpha _{\left(
\sigma _{\omega }^{J}\right) ^{2}2}^{\ast }+\frac{1}{2}\sum_{i=2}^{T}\left(
j_{i\Delta }^{\omega }-\mu _{\omega }^{JP}\right) ^{2}\right)
\end{eqnarray*}
\end{itemize}

\clearpage

\begin{table}[tbp]
\caption{Summary Statistics}
\label{Summary Statistics}
{\scriptsize
\flushleft{This table provides summary statistics for VIX and
VVIX index from January 3, 2007, to November 26, 2014.}}
\par
\begin{center}
{\scriptsize
\begin{tabular}{@{}crcccccrrrrrrrrr}
\toprule & \phantom{ab} & Mean & \phantom{ab} & Volatility & \phantom{ab} &
Skewness & \phantom{ab} & Kurtosis & \phantom{ab} & Min & \phantom{ab} & Max
& \phantom{ab} &  &  \\
\hline
\multicolumn{1}{l}{\ \ \ \ \ VIX} &  & \multicolumn{1}{r}{21.9101} &  &
\multicolumn{1}{r}{10.3966} &  & \multicolumn{1}{r}{2.1241} &  & 5.7181 &  &
9.89 &  & 80.86 &  &  &  \\
\multicolumn{1}{r}{VVIX} &  & \multicolumn{1}{r}{85.9204} &  &
\multicolumn{1}{r}{12.8226} &  & \multicolumn{1}{r}{0.8289} &  & 1.0079 &  &
59.74 &  & 145.12 &  &  &  \\
\bottomrule &  & \multicolumn{1}{r}{} &  & \multicolumn{1}{r}{} &  &
\multicolumn{1}{r}{} &  &  &  &  &  &  &  &  &
\end{tabular}
}
\end{center}
\par
\end{table}

\begin{table}[tbp]
\caption{VIX Parameter Estimates}
\label{parameter estimation result}
{\scriptsize
\flushleft{This table shows the parameter estimation results
for the four models using VIX and VVIX index data from January 3, 2007 to November 26, 2014. Four each parameter, we give the mean and the standard deviation of the posterior. "SV" denotes diffusion model with no jumps. "SVJ" introduces jumps in VIX in the SV model with constant jump intensity, "SVJJ-C" adds double jumps in VIX and its volatility in the SV model with constant jump intensity. "SVJJ-S" assumes the jump intensity to be stochastic in the SVJJ-C model.}%
}
\par
\begin{center}
{\scriptsize
\begin{tabular}{@{}crrrcrrcrrrrr}
\toprule & \phantom{ab} & \multicolumn{2}{c}{SV} & \phantom{ab} &
\multicolumn{2}{c}{SVJ-C} & \phantom{ab} & \multicolumn{2}{c}{SVJJ-C} & %
\phantom{ab} & \multicolumn{2}{c}{SVJJ-S} \\
\cmidrule{3-4} \cmidrule{6-7} \cmidrule{9-10}\cmidrule{12-13} &  & Mean &
Stddev &  & Mean & Stddev &  & Mean & Stddev &  & Mean & Stddev \\
\multicolumn{1}{l}{$\ \ \ \ \ \ \ \ \ \kappa _{V}$} &  & 1.6800 & 0.5733 &
& 1.5765 & 0.5600 &  & 1.8611 & 0.5688 &  & 2.1093 & 0.5866 \\
\multicolumn{1}{r}{$\varsigma _{V}$} &  & -1.1869 & 0.7718 &  & -0.8046 &
0.7673 &  & -0.2702 & 0.8305 &  & -0.1538 & 0.7820 \\
\multicolumn{1}{r}{$\theta $} &  & 2.3500 & 0.4404 &  & 2.3090 & 0.4446 &  &
2.2704 & 0.3954 &  & 2.3312 & 0.3120 \\
\multicolumn{1}{r}{$\kappa _{\omega }^{P}$} &  & 4.5162 & 1.0284 &  & 4.4308
& 0.9973 &  & 6.1132 & 1.0650 &  & 6.2849 & 1.0645 \\
\multicolumn{1}{r}{$\kappa _{\omega }^{Q}$} &  & 7.5104 & 0.4314 &  & 7.6866
& 0.4676 &  & 2.5996 & 0.2584 &  & 2.5674 & 0.1958 \\
\multicolumn{1}{r}{$\alpha _{\omega }$} &  & 3.8549 & 0.7807 &  & 3.7683 &
0.7540 &  & 4.0781 & 0.7882 &  & 3.7938 & 0.7308 \\
\multicolumn{1}{l}{$\ \ \ \ \ \ \ \ \ \rho $} &  & 0.5392 & 0.0161 &  &
0.5596 & 0.0141 &  & 0.5204 & 0.0169 &  & 0.4998 & 0.0190 \\
\multicolumn{1}{r}{$\sigma _{\omega }$} &  & 0.8560 & 0.0724 &  & 0.8207 &
0.0117 &  & 0.8848 & 0.0853 &  & 0.8461 & 0.0372 \\
\multicolumn{1}{r}{$\lambda _{0}$} &  &  &  &  & 0.6550 & 0.0492 &  & 2.4295
& 0.1787 &  & 2.7557 & 0.1332 \\
\multicolumn{1}{r}{$\lambda _{1}$} &  &  &  &  &  &  &  &  &  &  & 1.6086 &
0.1262 \\
\multicolumn{1}{r}{$\mu _{y}^{JP}$} &  &  &  &  & 0.1593 & 0.0220 &  & 0.1999
& 0.0279 &  & 0.1551 & 0.0171 \\
\multicolumn{1}{r}{$\mu _{y}$} &  &  &  &  & -0.0520 & 0.0037 &  & -0.0556 &
0.0746 &  & -0.0960 & 0.0306 \\
\multicolumn{1}{l}{$\ \ \ \ \ \ \ \ \ \sigma _{y}^{J}$} &  &  &  &  & 0.1075
& 0.0172 &  & 0.1121 & 0.0132 &  & 0.1231 & 0.0108 \\
\multicolumn{1}{r}{$\mu _{\omega }^{JP}$} &  &  &  &  &  &  &  & 0.1872 &
0.0226 &  & 0.1430 & 0.0239 \\
\multicolumn{1}{r}{$\mu _{\omega }$} &  &  &  &  &  &  &  & -2.0084 & 0.0882
&  & -1.2046 & 0.0547 \\
\multicolumn{1}{r}{$\sigma _{\omega }^{J}$} &  &  &  &  &  &  &  & 0.1307 &
0.0165 &  & 0.1420 & 0.0161 \\
\multicolumn{1}{r}{$\sigma _{P}$} &  & 0.0599 & 0.0077 &  & 0.0592 & 0.0071
&  & 0.0563 & 0.0082 &  & 0.0612 & 0.0076 \\
\bottomrule &  &  &  &  &  &  &  &  &  &  &  &
\end{tabular}
}
\end{center}
\par
\end{table}

\clearpage

\begin{table}[tbp]
\caption{Simulation results}
\label{Simulation results}
{\scriptsize
\flushleft{This table reports the $p$-values calculated by (\ref{p-value}) for all the
statistics of simulation results of VIX for different models. It describes the average comparisons of the statistics of historical data and the simulation paths from every given model. Very high or low p-values indicate the model's inability to capture the VIX dynamics. "SV" denotes diffusion model with no jumps. "SVJ" introduces jumps in VIX in the SV model with constant jump intensity, "SVJJ-C" adds double jumps in VIX and its volatility in the SV model with constant jump intensity. "SVJJ-S" assumes the jump intensity to be stochastic in the SVJJ-C model.}%
}
\par
\begin{center}
{\scriptsize
\begin{tabular}{@{}crcccccrrrr}
\toprule & \phantom{ab} & Data & \phantom{ab} & SV & \phantom{ab} & SVJ-C & %
\phantom{ab} & SVJJ-C & \phantom{ab} & SVJJ-S \\
\hline
\multicolumn{1}{l}{\ \ \ \ \ \ \ stadev} &  & \multicolumn{1}{r}{0.4493} &
& \multicolumn{1}{r}{0.0670} &  & \multicolumn{1}{r}{0.3351} &  & 0.8593 &
& 0.3733 \\
\multicolumn{1}{r}{skewness} &  & \multicolumn{1}{r}{0.9005} &  &
\multicolumn{1}{r}{0.1462} &  & \multicolumn{1}{r}{0.0359} &  & 0.2194 &  &
0.7479 \\
\multicolumn{1}{r}{kurtosis} &  & \multicolumn{1}{r}{0.7806} &  &
\multicolumn{1}{r}{0.2453} &  & \multicolumn{1}{r}{0.0150} &  & 0.6250 &  &
0.7186 \\
\multicolumn{1}{r}{maximum} &  & \multicolumn{1}{r}{4.7558} &  &
\multicolumn{1}{r}{0.0079} &  & \multicolumn{1}{r}{0.1383} &  & 0.6875 &  &
0.4744 \\
\multicolumn{1}{r}{minimum} &  & \multicolumn{1}{r}{2.3984} &  &
\multicolumn{1}{r}{0.2418} &  & \multicolumn{1}{r}{0.6040} &  & 0.0769 &  &
0.6896 \\
\multicolumn{1}{r}{maxjump} &  & \multicolumn{1}{r}{0.2267} &  &
\multicolumn{1}{r}{0.4897} &  & \multicolumn{1}{r}{0.1016} &  & 0.2805 &  &
0.6358 \\
\multicolumn{1}{l}{\ \ \ \ \ \ minjump} &  & \multicolumn{1}{r}{-0.2422} &
& \multicolumn{1}{r}{0.0953} &  & \multicolumn{1}{r}{0.9539} &  & 0.7698 &
& 0.7509 \\
\multicolumn{1}{r}{avgmax10} &  & \multicolumn{1}{r}{0.1852} &  &
\multicolumn{1}{r}{0.5697} &  & \multicolumn{1}{r}{0.1508} &  & 0.4308 &  &
0.2614 \\
\multicolumn{1}{r}{avgmin10} &  & \multicolumn{1}{r}{-0.1671} &  &
\multicolumn{1}{r}{0.3262} &  & \multicolumn{1}{r}{0.9069} &  & 0.7996 &  &
0.5826 \\
\multicolumn{1}{r}{perc0.01} &  & \multicolumn{1}{r}{-0.1345} &  &
\multicolumn{1}{r}{0.4271} &  & \multicolumn{1}{r}{0.8488} &  & 0.5005 &  &
0.6311 \\
\multicolumn{1}{r}{perc0.05} &  & \multicolumn{1}{r}{-0.0931} &  &
\multicolumn{1}{r}{0.4504} &  & \multicolumn{1}{r}{0.4957} &  & 0.5048 &  &
0.6277 \\
\multicolumn{1}{r}{perc0.95} &  & \multicolumn{1}{r}{0.0928} &  &
\multicolumn{1}{r}{0.5604} &  & \multicolumn{1}{r}{0.6182} &  & 0.8245 &  &
0.4581 \\
\multicolumn{1}{l}{\ \ \ \ \ \ perc0.99} &  & \multicolumn{1}{r}{0.1370} &
& \multicolumn{1}{r}{0.4478} &  & \multicolumn{1}{r}{0.0264} &  & 0.9379 &
& 0.1622 \\
\bottomrule &  & \multicolumn{1}{r}{} &  & \multicolumn{1}{r}{} &  &
\multicolumn{1}{r}{} &  &  &  &
\end{tabular}
}
\end{center}
\par
\end{table}

\clearpage

\clearpage

\begin{figure}[]
\par
\begin{center}
{\ \centering
\includegraphics[scale=0.5,bb=0 0 800 480]{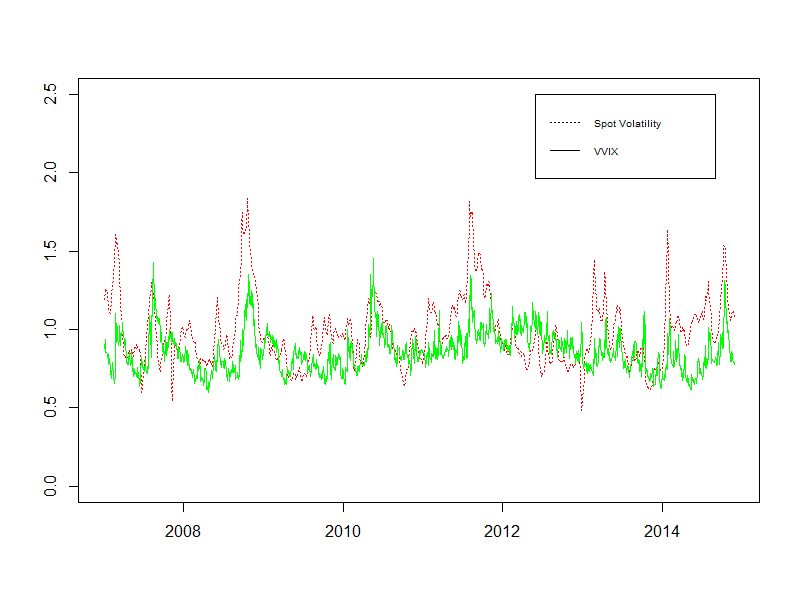} }
\end{center}
\par
{\scriptsize
\flushleft{The spot volatility in this figure is the estimated
posterior volatility of logVIX in SVJ model with only VIX index as the data
source. It shows the comparison of this volatility and the contemporaneous VVIX index}%
}
\caption{Spot Volatility from VIX estimation vs VVIX}
\label{Spot Volatility-VVIX}
\end{figure}

\clearpage

\clearpage

\clearpage

\clearpage

\clearpage

\clearpage

\clearpage

\normalsize
\end{document}